\documentclass[a4paper,10pt]{article}
\pdfoutput=1
\usepackage[utf8]{inputenc}
\usepackage{acronym}
\usepackage{lineno}
\usepackage{graphicx}
\usepackage{amsmath}
\usepackage{color}
\usepackage{hyperref}

\newcommand{\Pacc}{P^\mathrm{(acc)}}
\newcommand{\Pacci}{\Pacc_i}

\newcommand{\Prej}{P^\mathrm{(rej)}}

\newcommand{\traps}{\mathrm{traps}}

\newcommand{\rhoexp}{\rho_\mathrm{exp}}
\newcommand{\rhogauss}{\rho_\mathrm{gauss}}
\newcommand{\psiexp}{\psi_\mathrm{exp}}
\newcommand{\psigauss}{\psi_\mathrm{gauss}}
\newcommand{\Tc}{{T_\mathrm{c}}}
\newcommand{\bc}{{\beta_\mathrm{c}}}
\newcommand{\tw}{{t_\mathrm{w}}}
\newcommand{\eth}{{E_\mathrm{th}}}
\newcommand{\MAX}{\mathrm{max}}
\newcommand{\MIN}{\mathrm{min}}
\newcommand{\var}{\mathrm{var}}
\newcommand{\nr}{\mathrm{nr}}
\newcommand{\basin}{\mathrm{basin}}
\newcommand{\conf}{\mathrm{conf}}

\acrodef{rem}[REM]{Random Energy model}
\acrodef{erem}[EREM]{Exponential Random Energy model}
\acrodef{tm}[TM]{Trap model}
\acrodef{gtm}[GTM]{Gaussian Trap model}
\acrodef{mf}[MF]{mean field}
\acrodef{mc}[MC]{Monte Carlo}
\acrodef{iid}[i.i.d.]{independent identically distributed}

\graphicspath{{./FIGURE/}{./}{../plots/}}
\begin{document}

\begin{center}{\Large \textbf{
Activated Aging Dynamics and Effective Trap Model Description in the Random Energy Model
}}\end{center}

\begin{center}
M. Baity-Jesi\textsuperscript{1,2*}, 
G. Biroli\textsuperscript{1,3}, 
C. Cammarota\textsuperscript{4}
\end{center}

\begin{center}
{\bf 1} Institut de Physique Théorique, Université Paris Saclay, CEA, CNRS, F-91191 Gif-sur-Yvette, France
\\
{\bf 2} Department of Chemistry, Columbia University, New York, NY 10027, USA
\\
{\bf 3} Laboratoire de Physique Statistique, \'Ecole Normale Sup\'erieure, CNRS, PSL Research University, Sorbonne Universit\'es, 75005 Paris, France
\\
{\bf 4} King’s College London, Department of Mathematics, Strand, London WC2R 2LS, United Kingdom
\\
* mb4399@columbia.edu
\end{center}

\begin{center}
\today
\end{center}


\section*{Abstract}
{\bf 
We study the out-of-equilibrium aging dynamics of the Random Energy Model (REM) ruled by a single spin-flip Metropolis dynamics. We focus on the dynamical evolution taking 
place on time-scales diverging with the system size. Our aim is to show to what extent the activated dynamics displayed by the REM can be described in terms of an effective trap model. 
We identify two time regimes: the first one corresponds to the process of escaping from a basin in the energy landscape and to the subsequent exploration of high energy configurations, whereas the second one corresponds to the evolution from a deep basin to the other. By combining numerical simulations with analytical arguments we show 
why the trap model description does not hold in the former but becomes exact in the second.
}

\vspace{10pt}
\noindent\rule{\textwidth}{1pt}
\tableofcontents
\noindent\rule{\textwidth}{1pt}
\vspace{10pt}

\section{Introduction}
A large variety of liquids, if kept in the super-cooled metastable phase below the melting point, exhibit a spectacular steep increase of the relaxation time as the temperature decreases.
Despite the inherently disordered microscopic structure, this slowing down of the dynamics results in an effectively solid behaviour of the sample, which is commonly denominated glass \cite{debenedetti:01}.
At high temperatures, the system behaves as a simple liquid characterised by a fast dynamics \cite{berthier:11}.
When the temperature $T$ is lowered, motion both in real and configuration space becomes gradually slower
and the relaxation time extracted from dynamical correlation functions increases faster than Arrhenius in the so-called fragile liquids. Such behaviour is characteristic of activated 
dynamics in which a system in order to relax (and flow) has to jump over larger and larger barriers as the temperature is lowered, thus leading to a time-scale  
\begin{equation}\label{eq:arrhenius}
 \tau(\Delta E) = \tau_0 \exp(\Delta E/k_\mathrm{B} T)\,,
\end{equation}
where $\Delta E$ increases when $T$ decreases ($\tau_0$ sets the microscopic timescale, $k_\mathrm{B}$ is the Boltzmann constant which in this paper we will set to one).\\
A general theoretical understanding of activated dynamics, in particular approaching the glass transition, remains an open challenge although it is arguably the central problem of glassy dynamics, and it has implications and 
ramifications in many other fields from computer science to biology. There has been actually a lot of progress in the last decades in the study of infinite-dimensional or \ac{mf} glassy models 
\cite{castellani:05,charbonneau:17} both regarding thermodynamical and dynamical properties. However, since in \ac{mf} models the energy barriers are extensively high, 
activated dynamics takes place on time-scales diverging exponentially with $N$ and is very difficult 
to analyze; indeed so far 
analytical results have been obtained on the regime of times of order one  \cite{castellani:05} and  
there have been only few works which addressed the role and the determination of energy barriers \cite{cavagna:97,barrat:98,montanari:06,benarous:17}. 
New information has been recently gained from rigorous analysis 
of trap models and generalizations. As shown by Bouchaud \cite{bouchaud:92}, the \ac{tm} {\cite{dyre:87}} provides a simple framework to 
study activated dynamics.
The \ac{tm} describes a motion in the space of configurations in which the only way to reach a new configuration
is to attain a fixed energy threshold. From these high-energy threshold, it is possible to reach any other point of the phase space.
Time scales in the \ac{tm} are given by the Arrhenius law, and a detailed analysis of the dynamics can be
worked out {\cite{bouchaud:92,bouchaud:95,dyre:95,monthus:96}} (see \cite{denny:03,heuer:05} for comparison and applications to glassy liquids). In recent years these 
results obtained by physicists were shown to hold rigorously, and more than that, they were generalized in many other contexts \cite{benarous:06,benarous:02}. 
This line of research culminated in the proof that the \ac{rem} \cite{derrida:80} (the simplest mean-field glassy model)
endowed with single spin flip Metropolis dynamics has an aging dynamics on exponentially diverging time-scales effectively 
equivalent to the one of TM \cite{cerny:17,gayrard:17b,gayrard:16} \footnote{The proof of \cite{gayrard:17b} was made public recently, while this work was already at an advanced stage.}.
{Yet, it remained unclear \emph{how} \ac{tm} dynamics emerges in \ac{rem}. In this work we answer this open question by addressing the following points:  
(1) What happens at shorter, subexponential, time scales, 
(2) What is the mechanism that allows for a \ac{tm} description at exponentially large times, 
(3) How are observables influenced in the different time regimes,
(4) The role played by the energy landscape,
and (5) The importance of finite size effects. 
All this additional information will be instrumental in the task of further seeking for evidences of a trap-like dynamics in other disordered systems.}

The two key features that make the \ac{rem} different from the \ac{tm} are (i) that 
the states of the \ac{rem} are defined on an $N$-dimensional hypercube, instead of a fully connected space,
and (ii) that the system follows physical dynamics, such as single spin flip Metropolis \ac{mc}, instead of the purely
barrier-passing trap dynamics. \\
These differences make \ac{rem} dynamics more difficult and interesting than the one of \acp{tm}. In particular, 
it retains some of the essential ingredients of activated dynamics present in more realistic systems. Its study 
is therefore a first step to develop a general theory. 
Developing a general understanding of 
how and to what extent \ac{tm}-like dynamics emerge at long times in the \ac{rem} is the aim of our work. In order to do this, 
we study dynamics on exponentially growing time-scales by performing numerical simulations of finite systems. 
This approach was pioneered by Crisanti and Ritort \cite{crisanti:00} to understand properties of activated glassy 
dynamics. Previous results have been obtained for a modified (exponential) \ac{rem} using a 
microscopic Hamiltonian based on the number partitioning problem \cite{junier:04}. It was shown that 
many relevant observables such as the distribution of trapping times and
the aging functions (autocorrelation functions), are indeed trap-like. \\
In the following, after setting up the notations and recalling the main models and results used in this work, 
we shall first show to what extent and on which time-scales \ac{rem} dynamics is different from TM one. We will then
explain and show how on very large time-scales the TM dynamics emerge. 

\section{Models and simulations}
This section is devoted to recall the definition of the models we will refer to, list the results useful for our work, and set the notation. 
\subsection{Trap Models} \label{se:tm}
Trap Models (\ac{tm}) were proposed to give a first simplified framework for activated aging dynamics~\cite{bouchaud:92,dyre:95}.
They are defined in terms of a collection of $M$ configurations whose  
 energies $E_i$ are \ac{iid} random variables extracted from an exponential distribution
\begin{equation}\label{eq:distr-expo}
 \rhoexp(E) = \frac{1}{\alpha}\exp(E/\alpha) \Theta(-E)\ ,
\end{equation}
where $\alpha>0$ is the mean of the exponential distribution, and $\Theta(x)$ is the Heaviside step function; when $M=2^N$, these configurations may represent all the possible configurations of a system with $N$ binary degrees of freedom, e.g. Ising spins.\\
Their continuous time dynamics is defined as follows:
at each time step a configuration $i$ is randomly chosen with flat probability $1/M$ 
among all the configurations at disposal, the system remains trapped in the configuration $i$ 
for a \emph{trapping time} $t$
which is a random variable extracted from an exponential distribution with mean 
\begin{equation}\label{eq:trap-time}
\tau_i=\tau_0\exp(-\beta E_i) \ ,
\end{equation}
where $\beta=1/T$ is the inverse temperature, and $\tau_0$ sets the time unit. Once it moves, the system jumps to a configuration drawn uniformly from the $M$ available ones and the process described above repeats. \\
Note that the inverse of the average trapping time can be interpreted as the (Arrhenius-like) probability for the system to escape from its trapping configuration $i$
\begin{equation}
 P_{Arr}= \exp(-\beta \Delta_i)\,,
\end{equation}
once the configuration is surrounded in all the directions by barriers of height $\Delta_i=E_{top}-E_i$ with $E_{top}=0$.
In other words, the energy landscape is like a golf-course: during the dynamics the system needs to jump up to a fixed top level of energy 
$E_{top}$ in order to access the rest of the energy landscape.  
Once this is achieved, the dynamics looses completely memory of the past because any configuration is  accessible from any other irrespectively from the past dynamical history, and any new configuration is equally acceptable.
Due to these two features \ac{tm} dynamics is a \emph{renewal process} \cite{cox:62} because after every jump the process starts anew. Despite this strong simplification, the dynamics retains a number of features that lead to a non trivial aging phenomenology, as it will be evident in what follows.\\
It can be easily shown that the mean trapping times $\tau$ obtained from trapping configurations with random energies $E$ are random variables, power-law distributed \cite{bouchaud:92}
\begin{equation}\label{eq:ptau-expo}
 \psiexp(\tau) = \frac{T}{\alpha\tau}\left(\frac{\tau}{\tau_0}\right)^{-T/\alpha}\Theta(\tau-1)\propto\frac{1}{\tau^{1+T/\alpha}}\,\,,\,\tau\gg1.
\end{equation}
The regime $T/\alpha<1$ (or $T<\alpha=T_c$) corresponds to a dynamics 
with several interesting properties \cite{bouchaud:92}. 
Among these, the configurational average of the mean trapping times $\tau_i$ is infinite (correspondingly 
the thermodynamic limit is not defined: the partition function is divergent).
Moreover taking a collection of $n$ of these random variables $\{\tau_i\}$ their sum is dominated by the maximum among them: $\sum_i^n\tau_i\sim\tau_{\MAX}$.
Equivalently, at a finite time from the beginning of the dynamics,  $t_w=\sum_i^n\tau_i$, the system is trapped with finite probability in the lowest trap among the ones already explored (the initial condition is drawn uniformly from 
the $M$ configurations). 
The energy $E_{\MIN}$ of the lowest trap can be therefore  evaluated equating its trapping time  $\tau_{\MAX}\simeq\tau_0\exp(-\beta E_{\MIN})$ to $t_w$ and inverting the relation. Given that a great fraction of time is spent 
in the lowest trap, the average energy also scales similarly with $t_w$: 
\begin{equation}\label{eq:et}
 E(t_w)= -T\log\left(\frac{t_w}{\tau_0}\right)\,.
\end{equation}
This is an example of aging behaviour in which the relaxation time of the system is set by its age \cite{cugliandolo:03,biroli:16b}. \\
\ac{tm} are so simple that it is possible to explicitly evaluate average correlation functions along the dynamics.
The dynamical observable considered in the literature is the persistence function $C(t_w,t_w+t)$: it is non zero (and equal to $1$) as long as the system remains trapped in the same configuration between $t_w$ and $t_w+t$. As 
soon as the system escapes from the trapping configuration it occupied at $t_w$, the persistence drops to zero.
The average value of the persistence is given by the probability of not changing configuration between the two observed times $t_w$ and $t_w+t$, which is traditionally defined as $\Pi(t_w,t_w+t)$.
This quantity has been found~\cite{bouchaud:92} in the large time limit to only depend on the ratio $w=t/t_w$ according to the so-called Arcsin law~\cite{bouchaud:95,benarous:06}: 
\begin{equation}\label{eq:pi}
\lim_{t_w\rightarrow\infty,t/t_w=w} \Pi(t_w,t_w+t)= \frac{\sin(\pi x)}{\pi}\int_{w}^{\infty}\frac{du}{u^x(1+u)} \equiv H_x(w) \ ,
\end{equation}
where the parameter $x$ depends on the mean $\alpha$ of the exponential distribution $\rhoexp(E)$, and on the temperature $T$ of the dynamics as $x=T/\alpha<1$. \\

\paragraph{Gaussian Trap Model}
We will want to compare the \ac{tm} to the \ac{rem}, defined in the next section. 
With this objective, as a first minor step towards a more realistic model for activated aging dynamics, we also take into account the \ac{gtm},
in which the distribution of the energy is Gaussian, 
\begin{equation}\label{eq:distr-gauss}
\rhogauss(E)=\frac{1}{\sqrt{2\pi N}}\exp\left(-\frac{E^2}{2 N}\right) \ ,
\end{equation}
and the distribution of trapping times is therefore
\begin{equation}\label{eq:ptau-gauss}
 \psigauss(\tau) = \frac{T}{\tau\sqrt{2\pi N}}\exp{\left({-\frac{T^2}{2 N}\log(\tau/\tau_0)^2}\right)}\,.
\end{equation}
Note that $\psi(\tau)$ is in this case lognormal. 
\\
This last model is directly inspired by the \ac{rem}, introduced in~\cite{derrida:80}: the equilibrium properties of \ac{gtm} (with $M=2^N$) and \ac{rem} coincide, but 
the dynamical rules are different. In the case of the \ac{gtm}, these are equivalent to the ones defined before (for the \ac{rem} see next section). 
From equilibrium studies, it is known that this model shows a thermodynamic entropy-vanishing transition at a critical temperature $\Tc\equiv1/\bc=\frac{1}{\sqrt{2\log(2)}}\simeq0.849$ \cite{derrida:80}.
For what concerns the dynamics, the \ac{gtm} presents a dynamics with an infinite number of time-scales. 
On each one of them the dynamics is effectively identical to the one of an exponential trap model with a 
time-scale dependent $\alpha$ \cite{benarous:08}. More precisely, the continuous time dynamics of the \ac{gtm} behaves like the one of a \ac{tm} with a parameter 
$\alpha(t_w)=T/x(t_w)= N/(T\log t_w)<1$ dependent on the age of the system~\cite{monthus:96} if the times at which the dynamics is observed are properly 
rescaled~\cite{bouchaud:95,monthus:96,cammarota:17}.\footnote{This becomes evident if we rewrite \eqref{eq:ptau-gauss} in the form of an exponential distribution in analogy with \eqref{eq:ptau-expo} 
and define an effective $T/\alpha\sim (T^2\log\tau)/N$ where we know that the relevant time scale of the system $\tau$ is given by its age.}

\subsection{Random Energy Model}
As anticipated in the previous section, following its classical definition~\cite{derrida:80} we call \ac{rem} a model that is defined as a collection 
of $M=2^N$ configurations representing all the possible microstates of a system with $N$ binary variables, that we call spins.
To each configuration is associated an \ac{iid} quenched random variable extracted from the Gaussian probability distribution
$\rhogauss(E)$, previously shown in equation \eqref{eq:distr-gauss}.
\footnote{The variance of $\rhogauss(E)$ is slightly different from Derrida's original definition \cite{derrida:80}, 
since we have $\var{E}=N$ instead of $N/2$. This conveys that the critical temperature is rescaled by a factor $\sqrt{2}$.}\\
The most natural choice of dynamics to be considered for such a system is a single spin flip \ac{mc} Metropolis dynamics \cite{krauth:06}, the study of which has been the motivation of a number of previous rigorous studies~\cite{gayrard:16,gayrard:17b} and will be the focus of this work.\\
Beside the classical \ac{rem}, to fill the gap between the \ac{rem} and \ac{tm}, we will also consider a slight modification of this model that we will call \ac{erem} 
\cite{bouchaud:97}. In this model the distribution of energies is exponential $\rho(E)=\rho_{\exp}(E)$; also for this model we consider a single spin flip Metropolis dynamics as we do for the \ac{rem}.\\
We underline again that from the {\it equilibrium} point of view \ac{rem} and \ac{erem} are alike \ac{gtm} and \ac{tm}, respectively.
Dynamically, two major elements of difference are introduced: (1) the system is not in general expected at every step to go back to the highest possible energy level at disposal, $E_\mathrm{top}$; 
(2) since only single spin flip moves are considered the number of neighbouring configuration is $N$ (instead of $2^N$), hence the number of dynamical paths at disposal is considerably reduced. 
In this case, the lattice representing the dynamical connections between configurations separated by a single move is not fully connected. It is instead a hypercube in dimensions $N$ with edges 
corresponding to the moves induced by the spin-flips.  
These two new ingredients, despite the absence of static correlation between the energies of any pair of configurations, might introduce correlations in the energies sampled during the dynamics, and considerably change the aging behaviour.\\
Note that these two ingredients must be taken into account if we aim at develop the description of an out-of-equilibrium dynamics for more realistic systems where dynamics is always controlled by local moves and typical dynamical paths may allow relaxation without stepping through high energy levels.
The additional introduction of static correlations between the energies of neighbouring configurations is a further element to consider when developing a theory of activated dynamics of realistic systems, but it will not be the argument of this work.
\\
\subsubsection{Basic Features of the Energy Landscape}
\begin{figure}[!htb]\centering
 \includegraphics[width=0.8\columnwidth]{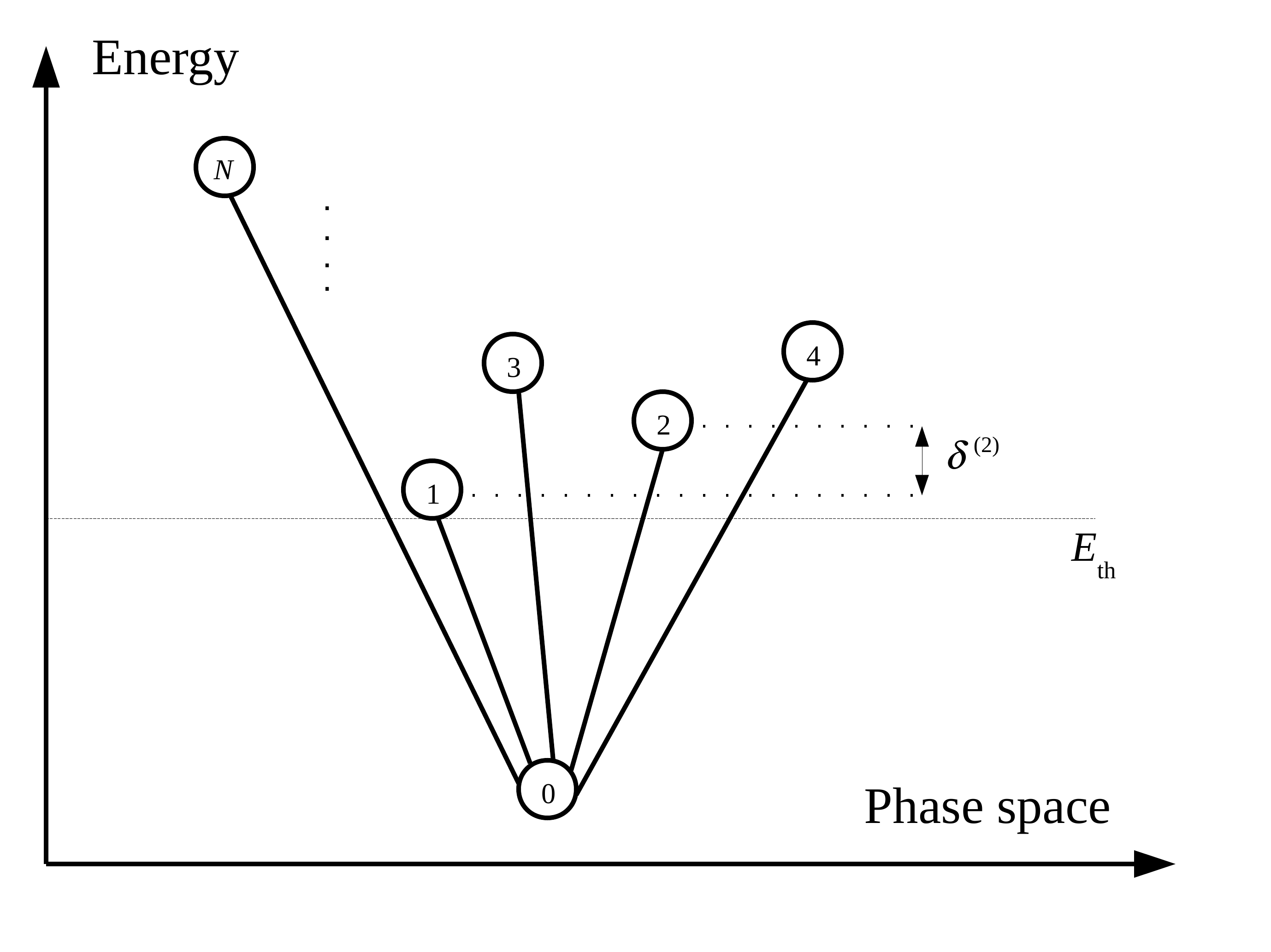}
 \caption{Schematic diagram of the connectivity of a low-lying configuration $C_0$, with its neighbours $C_1, C_2,\ldots,C_N$. $\eth$ is the threshold energy, below which we say that the system is in a basin. We
 define $\eth$ as the typical value of the smallest neighbour of any configuration (i.e. $\eth=E^{(1)}$). The energy difference $\delta^{(2)}$ 
 is defined in Eq. \eqref{eq:kth-deepest-diff}.}
 \label{fig:diag-cc}
\end{figure}

Before focusing on the out of equilibrium dynamics of these two models we first list a number of features that characterise the typical local arrangement of energy levels around a given (say deep) configuration $C_0$. \\
In order to do this, we shall use results on extreme value statistics \cite{galambos:87} that we recall by using simple scaling arguments. We evaluate the typical energy level of the lowest configuration, 
$C_1$, among the $N$ configurations surrounding a given $C_0$, of the second lowest configuration, $C_2$, and so on and so forth, as it is represented in Fig.~\ref{fig:diag-cc}.
Since the energies of configurations are uncorrelated, these typical energy levels will not depend on $C_0$. 
The energy $E^{(1)}$ of the lowest configuration among the nearest neighbours of $C_0$ is the minimum of $N$ Gaussian variables, whose scaling can be obtained by evaluating the probability that 
out of the $N$ energies at disposal one has an energy contained in the interval $[-\infty,E^{(1)}]$. The energy scale 
where this probability starts to be of the order of one sets the scale of $E^{(1)}$, leading to the equation:  
\begin{equation}\label{eq:kth-deepest}
 O(1)\sim N\int_{-\infty}^{E^{(1)}}dE \rho(E) \ .
\end{equation}
In the \ac{rem} this leads to 
\begin{equation}\label{eq:kth-deepest-rem0}
 E^{(1)}\sim -\sqrt{N\log N}+x_1\sqrt{\frac{N}{\log N}} \ ,
\end{equation}
where the probability law of the order one variable $x_1$ can be obtained from extreme value statistics (it is the derivative of a Gumbel law) \cite{galambos:87}. The result for the energy 
$E^{(k)}$ of the $k^\mathrm{th}$ lowest configuration among the first dynamical neighbours of $C_0$ is analogous: 
\begin{equation}\label{eq:kth-deepest-rem}
 E^{(k)}\sim -\sqrt{N\log N}+x_k\sqrt{\frac{N}{\log N}} \ .
\end{equation}
where of course $x_1<x_k$ with probability one. 
The same arguments in the \ac{erem} lead to 
\begin{equation}\label{eq:kth-deepest-erem}
 E^{(k)}\sim -\alpha \log N+x_k\ .
\end{equation}
Note that the difference in energy, 
\begin{equation}\label{eq:kth-deepest-diff}
 \delta^{(k)} \sim E^{(k)}-E^{(1)}\,,
\end{equation}
between the $1^\mathrm{st}$ and the $k^\mathrm{th}$ deepest levels among $N$ configurations at disposal is diverging with $N\rightarrow \infty$ in the \ac{rem}, 
\begin{equation}\label{eq:kth-deepest-diffrem}
 \delta^{(k)} \sim \sqrt{\frac{N}{\log N}} \ ,
\end{equation}
and it is of order one in the \ac{erem}.\\
A final fundamental remark is that the ground state, or the configuration with lowest energy, $E_{GS}$, among all the $M$ at disposal can be obtained from the equation~\eqref{eq:kth-deepest} by substituting $N$ with $M$. It gives in the \ac{rem}
\begin{equation}\label{eq:min-rem}
 E_{GS}\sim -\sqrt{2N\log M}
\end{equation}
and in the \ac{erem}
\begin{equation}\label{eq:min-erem}
 E_{GS}\sim -\alpha \log M.
\end{equation}
When we consider a number of configuration $M=2^N$, the typical ground state energy is in both cases extensive and corresponds to $E_{GS}\sim -N\sqrt{2\log 2}$ in the first case, and to $E_{GS}\sim \alpha N \log 2$ in the second case.\\
Low temperature dynamics of \ac{rem} and \ac{erem} (with $T<1/\sqrt{2\log 2}$ and $T<\alpha$, respectively) will tend to reach this ground state energy in the long time limit, leading the system to be trapped for long time intervals in very 
deep configurations $C_0$ surrounded by $N$ configurations which are well above in the energy landscape (leading to a golf-course like landscape as stressed previously). 
The system has to progressively decrease its intensive energy; however any configuration with a negative intensive energy is surrounded by neighbours that have typically zero intensive energy.
This leads to an extremely slow and activated dynamics which proceeds on time-scales diverging exponentially with $N$. \\
An important difference between the two models is that, for the \ac{rem}, dynamics are activated and aging is observed until the system reaches equilibrium, for $T$ both over and under $\Tc$.
In the \ac{erem}, instead, only the low-temperature phase has interesting dynamics. For $T>T_c$ the dynamics of the \ac{erem} is fast and not interesting for our purposes since the system 
remains floating at zero intensive energy, and does not display aging.

\subsubsection{Energy Basin Description}
\begin{figure}[!ht]
\centering
 \includegraphics[width=.8\textwidth,trim={2cm 13cm 2cm 0cm}]{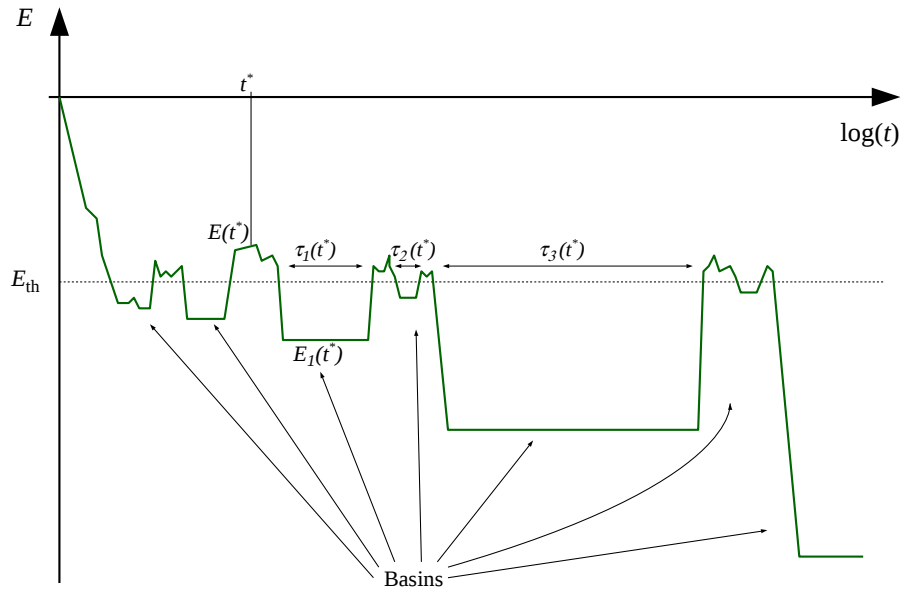}
 \caption{A sketch of the energy evolution in either the \ac{erem} or the \ac{rem}.
 At a given time $t^*$, the instantaneous energy is $E(t)$. Independently of whether at time $t^*$ the system is in a basin or on a barrier, 
 the energy of the next basin is $E_1(t^*)$, calculated via Eq.~\eqref{eq:ebasin}. The times $\tau_1(t^*),\tau_2(t^*)$ and $\tau_3(t^*)$ are the trapping times of the three basins following $t^*$.
 }
 \label{fig:diagramma-evol}
\end{figure}

In \ac{tm}s, configurations are, by definition, schematic approximations of the deep basins, that in more realistic models are made up of several configurations with correlated energies. 
At first sight also for the \ac{rem} and \ac{erem} basins consist in single configurations as we have seen in the previous section. However basins containing more than one configuration do exist (despite they are much rarer than basins with single configurations, as a matter of fact they are still exponentially numerous in $N$~\cite{gayrard:16}) and as we shall show they can be relevant for the dynamical behaviour, at least for certain observables. 
In the following, in order to test if a 
 \ac{tm} description holds for the dynamics of the  \ac{rem} and \ac{erem} we need to map the 
 \ac{mc} Metropolis dynamics into a sequence of basin jumps, and hence give a pragmatical definition of basins. 
This mapping can be used also in more realistic models; by studying it for the \ac{rem} and \ac{erem} we shall unveil subtleties and properties related to this mapping that can be useful when it is applied to more realistic cases.\\
Given the high dimensionality of the energy landscape, it is not straightforward to partition the system in a set of distinct basins.
Therefore, inspired by a similar analysis proposed recently in~\cite{cammarota:15}, we use a dynamical definition of basin, which focuses on the one-dimensional dynamical path described by the energy evolution as a function of time, $E(t_w)$.
We choose a threshold energy $\eth$ and say that each time $E$ becomes smaller than $\eth$ the system is \emph{in a basin}. When finally $E>\eth$ again, the basin is abandoned, and we say we are \emph{on the barrier} between two basins.
A schematic representation is provided in Fig.~\ref{fig:diagramma-evol}.
A natural choice for $\eth$ is the average minimum energy $E^{(1)}$ of the $N$ configurations that can be reached from any configuration $C_0$ in a single dynamical step. 
With this choice, almost all the neighbours of any configuration are typically at $E\geq\eth$.
This energy level does not depend on the configuration $C_0$ visited by the dynamics, whose energy $E_0$ is slowly decreasing towards $E_{GS}$. 
Hence the distance between $\eth$ and the energy of the configuration typically explored by the dynamics will become larger and larger in the long time limit in a similar way as in \ac{tm} dynamics. \\
Note that for short times, the visited basins are typically very close to $\eth$, implying that the system has to age enough time before
the energy basins become similar to traps, where the system remains stuck for long times.
In full generality, since within each basin $B$ the system overwhelmingly spends the most of the time
in its lowest-energy configuration, we associate to $B$ the energy of the deepest configuration that belongs to that basin: 
\begin{equation}\label{eq:ebasin}
E_\mathrm{basin} = \min_{C_i\in\text{B}}E(C_i)\,.
\end{equation}
The same applies to any static property we mention of a basin: it refers to the lowest-energy configuration within the considered basin.\\
Finally, to each basin we associate a trapping time as shown in figure \ref{fig:diagramma-evol}.

\section{Absence of trap-like behaviour on intermediate time scales}\label{sec:non-trap}
In this section we shall show that the dynamics of \ac{rem} and \ac{erem} is not trap-like 
on time-scales corresponding to the exit from a basin and the exploration of configurations 
at the threshold level. As we discuss in the next section, 
the mapping to the \ac{tm} only emerges on very large time-scales in a coarse-grained way.   \\
We first present numerical results concerning the dynamics of the \ac{rem} and \ac{erem}.
For each series of simulation, we focus on three temperatures in the 
glassy phase, $T=0.25, 0.50, 0.75$, and $T=1.50$ in the disordered phase. 
We remind the reader that $\Tc=\alpha$ (we choose $\alpha=1$) in the \ac{erem}, and $\Tc=\frac{1}{\sqrt{2\log(2)}}\simeq0.849$ in the \ac{rem}.\\
The system sizes are $N=8,12,16,20,24$. The length of the run is $t_\mathrm{run}=2^{1+A+N}$ \ac{mc} steps.
The value of $A$ was decided according to the wallclock time availability.
The amount of samples $N_\mathrm{sam}$ per parameter set, and other details of the simulations
are described in Table \ref{tab:sim}.
\begin{table}[!th]
\begin{center}
\begin{tabular}{ccccc}
$T$ & $N$ & $A$ & $t_\mathrm{run}$ (\acs{mc} steps) & $N_\mathrm{sam}$\\\hline\hline
0.25, 0.50, 0.75 &  8 & 18 & $10^8$ & 20000\\
0.25, 0.50, 0.75 & 12 & 18 & $10^9$ & 12000\\
0.25, 0.50, 0.75 & 16 & 18 & $10^{10}$ & 6000\\
0.25, 0.50, 0.75 & 20 & 18 & $10^{12}$ & 1500\\
0.25, 0.75        & 24 & 18 & $10^{13}$ &  200\\
1.50             & 16 & 0 & $10^5$ & 100\\
1.50             & 20 & 0 & $10^6$ & 100\\
1.50             & 24 & 0 & $10^7$ & 100
\end{tabular}
\end{center}
\caption{Parameters of the simulations. We performed the same simulations both for the Exponential and the Gaussian \ac{rem}. The exact number 
of \ac{mc} steps of each run is $t_\mathrm{run}=2^{1+A+N}$. In the column of $t_\mathrm{run}$ we display the order of magnitude of this number.
$N_\mathrm{sam}$ is the number of independent samples simulated with the given set of parameters.}
\label{tab:sim}
\end{table}

\subsection{Non-renewal behaviour}
In the following we focus on several observables that in the \ac{tm} show that the dynamics is a renewal process. 
This means roughly speaking that every time a basin is abandoned, the system should completely loose memory of its past. As we shall show, the behaviour we find is quite different.

\subsubsection{Energy of the next basin}

\begin{figure}[!htb]
\includegraphics[width=0.49\textwidth]{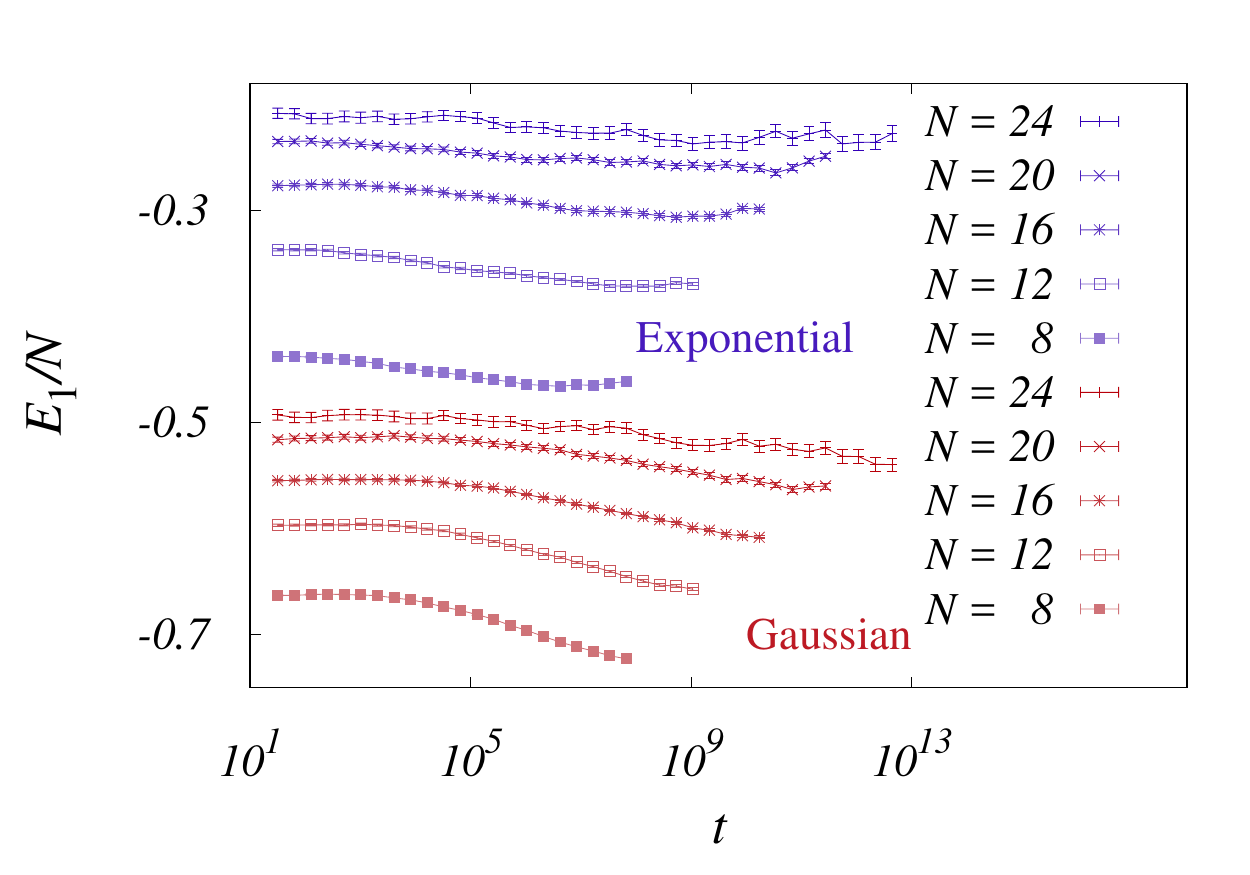}
\includegraphics[width=0.49\textwidth]{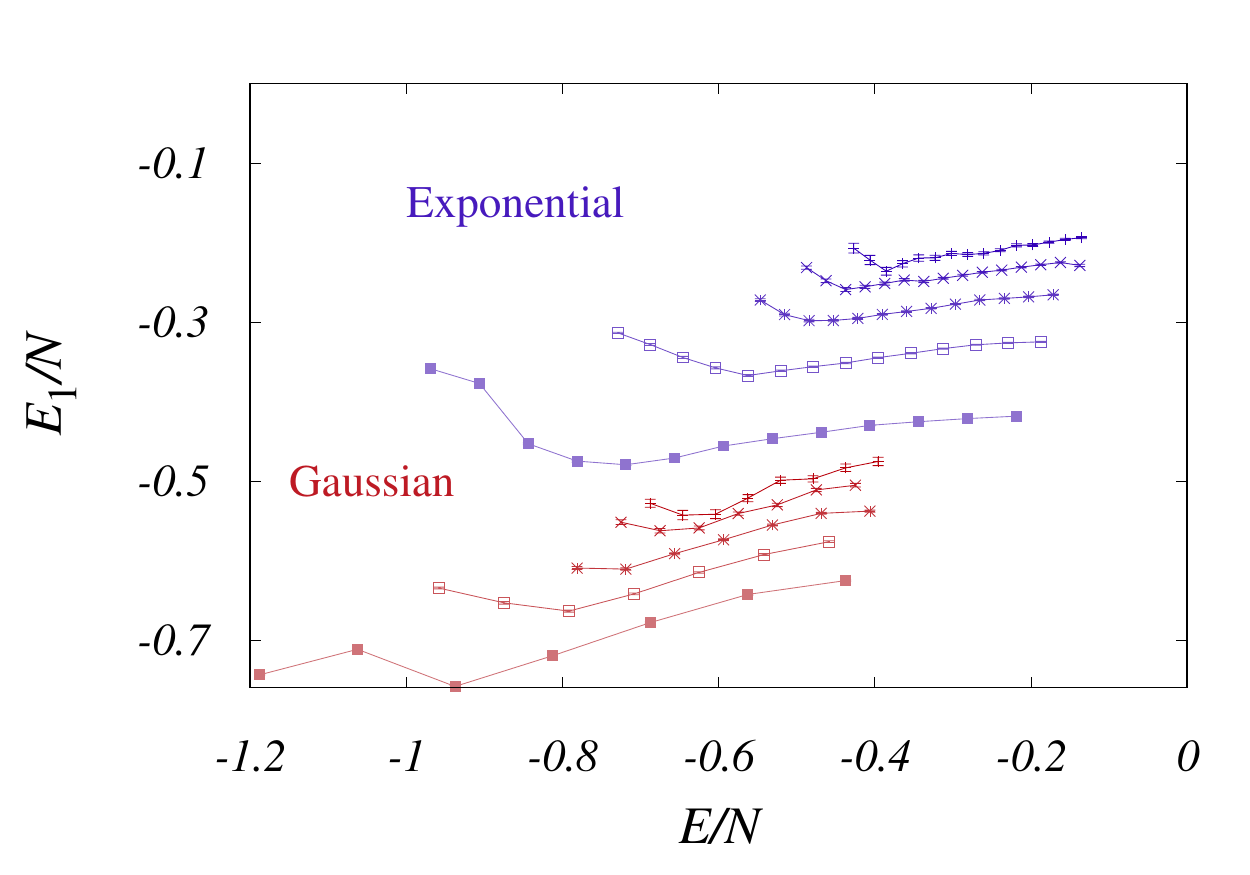}
\caption{Energy of the next basin, $E_1$, in the \ac{erem} (blue curves) and in the \ac{rem} (red curves), at $T=0.25$. On the 
left side we show it as a function of time, on the right side we show it as a function of the previous basin's energy. The data for the \ac{erem} 
levels out when the system has thermalised (thermalisation times can be deduced from figure \ref{fig:etbar}).
In the right plot, we removed the points at lowest energy because they were fluctuating.
See also the caption of figure \ref{fig:tau123} for remarks on the measurement protocol.}
\label{fig:eplus}
\end{figure}

If we fix a time $t$, the system will likely be in a basin. As shown in Sec.~\ref{sec:elogt}, the average energy $E(t)$ 
depends on $t$ approximatively as described by the relation \eqref{eq:et}. So, indeed, the system is trapped 
in deeper minima at long times as in the   \ac{tm}. However, in the \ac{tm}, once a basin is abandoned then the energy of the following one, $E_1(t)$ (see Fig.~\ref{fig:diagramma-evol}),
should be completely uncorrelated, i.e. $E_1$ is independent
of $E(t)$ (and hence of $t$). The situation is quite different in our simulations of the dynamics of \ac{rem} and \ac{erem} as shown in figure \ref{fig:eplus}, where $E_1(t)$ is instead clearly dependent on $t$ and $E(t)$.

\subsubsection{Trapping time of the following basins}
\begin{figure}[!htb]
\includegraphics[width=\textwidth]{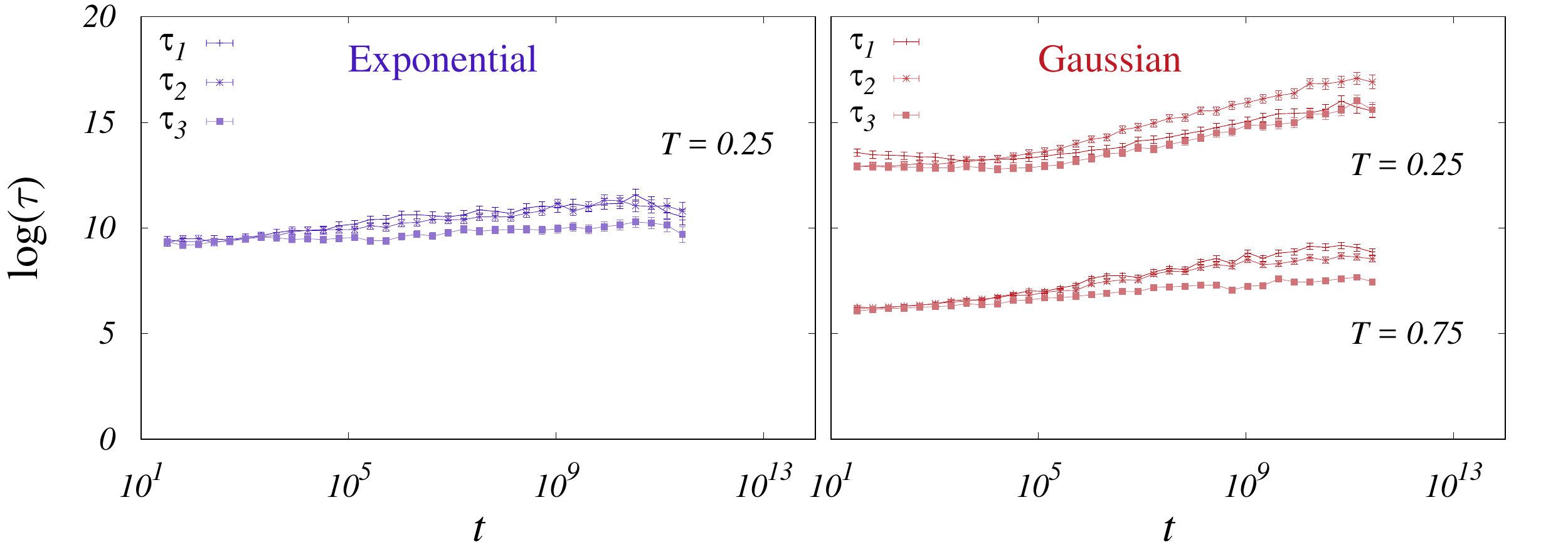}
 \caption{The trapping times $\tau_{1},\tau_{2}$ and $\tau_{3}$ in the \ac{erem} (left, $T=0.25$) and \ac{rem} (right, $T=0.25, 0.75$), for $N=20$.  In the \ac{erem}, results for $T=0.75$ are not reported, because no sizable growth of $\tau_1$, $\tau_2$ and $\tau_3$ was found. The time dependence is fairly visible, especially in the Gaussian case. The growth of $\tau_1$ is stronger than that of $\tau_2$ and $\tau_3$, since it is the trapping time that preserves the highest memory of $t$. 
Our measuring procedure excludes basins that, at the end of the run, have not been abandoned. This gives a bias in the final points of the curves, since towards the end of a run large basins are rejected more often than small ones. The growth we observe is clearly visible despite this bias.
 }
 \label{fig:tau123}
\end{figure}

As well as the energy of the next basin, also the statistics of the trapping time should be time-independent if the dynamics
is renewal. We call $\tau_1(t)$ the trapping time of the basin that immediately follows the one found at $t$, $\tau_2(t)$ the energy of the next basin, and $\tau_3$ the one of the 
third basin after $t$. A schematic description of the definition of $\tau_1,\tau_2$ and $\tau_3$ is provided in Fig.~\ref{fig:diagramma-evol}.\\
We see from figure \ref{fig:tau123} that $\tau_1,\tau_2$ and $\tau_3$ are not time-independent in the aging regime, again in contradiction with the renewal property of a trap-like dynamics.

\subsubsection{Rank of the deepest trap}
\begin{figure}[!htb]
\includegraphics[width=\textwidth]{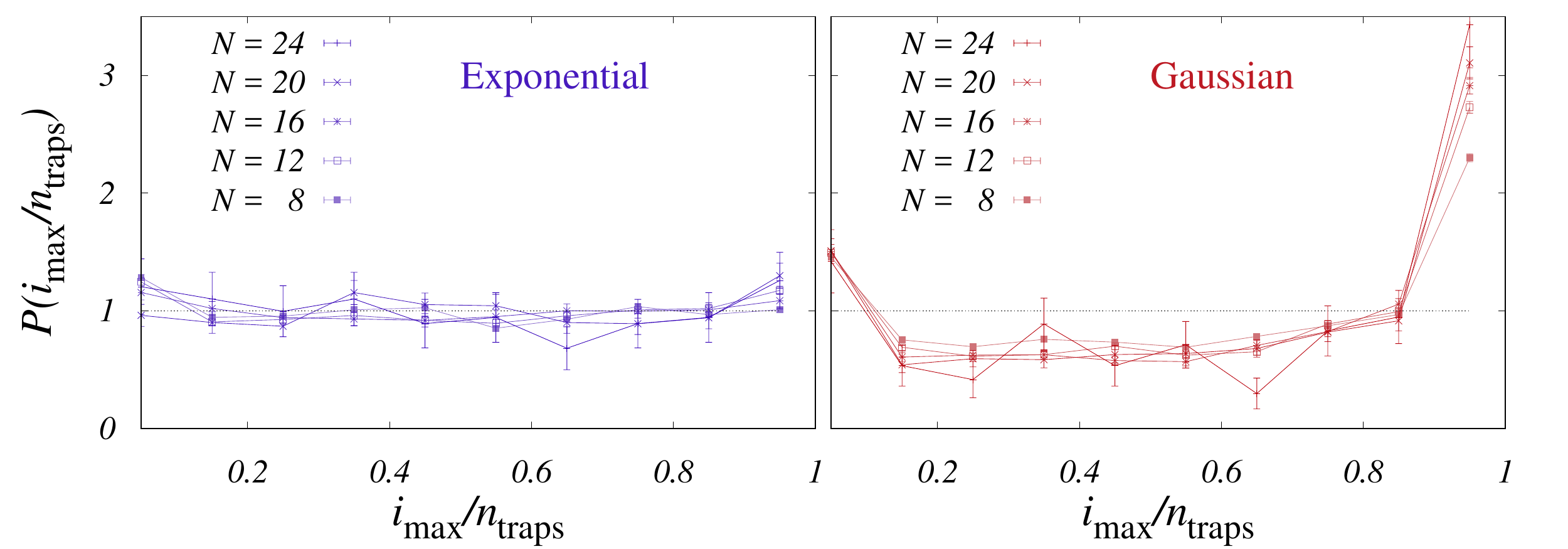}
 \caption{The distribution $P(i_\mathrm{max}/n_\mathrm{traps})$ in the Exponential (left) and Gaussian (right) \ac{rem}, at $T=0.25$.
 In the Gaussian \ac{rem}, $P(i_\mathrm{max}/n_\mathrm{traps})$ is significantly non-flat, in contrast with the \ac{tm} prediction.
 }
 \label{fig:imax}
\end{figure}

After $n_\traps$ basins have been visited (we only count the last basin if the run finished on a barrier), 
one can ask when the longest trap has been visited. To be more precise, we assign a rank $i$ to each basin, according to the order of occurrence. So, $i=0$ indicates the first basin,
$i=1$ the second basin, and so on, until $i=n_\traps-1$ for the last basin. 
If the dynamics is renewal, the trapping time $\tau_i$ of basin $i$ follows the same probability distribution for any $i$. 
Hence, any of the visited basins can be the one with the longest trapping time with equal probability.
Calling $i_\MAX$ the rank of the basin with the longest $\tau$, the distribution $P(i_\MAX/n_\traps)$ is flat in \acp{tm}.
In figure \ref{fig:imax} we show that for the \ac{rem} this is not the case.

\subsection{Back and forth in the same basin}
\begin{figure}[!htb]
\centering
\includegraphics[width=0.8\textwidth]{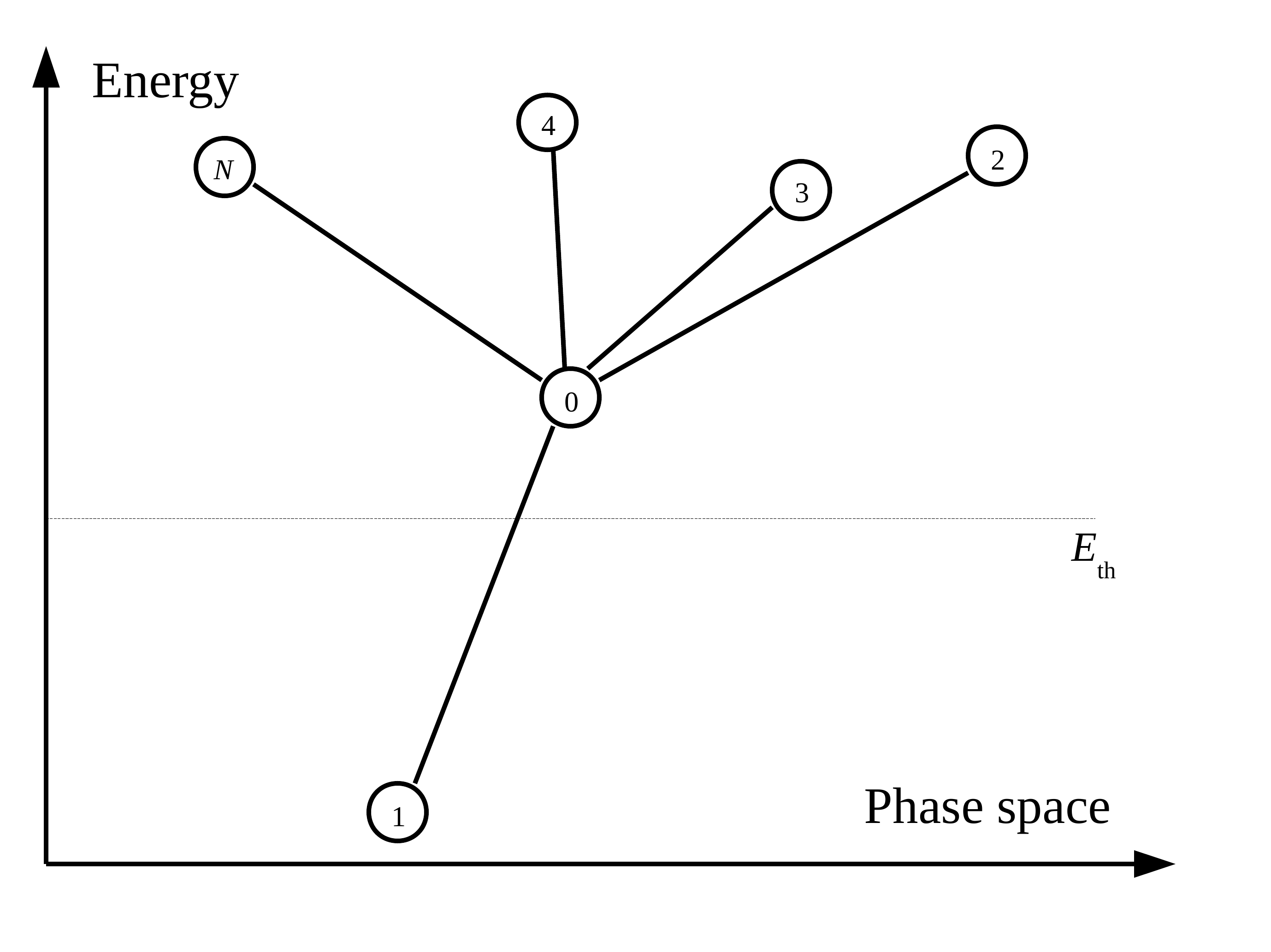}
 \caption{The diagram shows the situation we take in account for our computation. 
 Each circle indicates a configuration in phase space, with its energy. The 
 horizontal line marks the threshold energy $\eth$.
 We consider the case in which the system is found on the edge of a basin. It has just reached the configuration $C_0$  moving from $C_1$ which lies deep down in a basin. In order to go immediately back to $C_1$ the 
 subsequent $\mathcal{O}(N)$ moves towards its other $N-1$ neighbours need to be rejected.
 }
 \label{fig:diagram}
\end{figure}

\begin{figure}[!htb]
\includegraphics[width=0.49\textwidth]{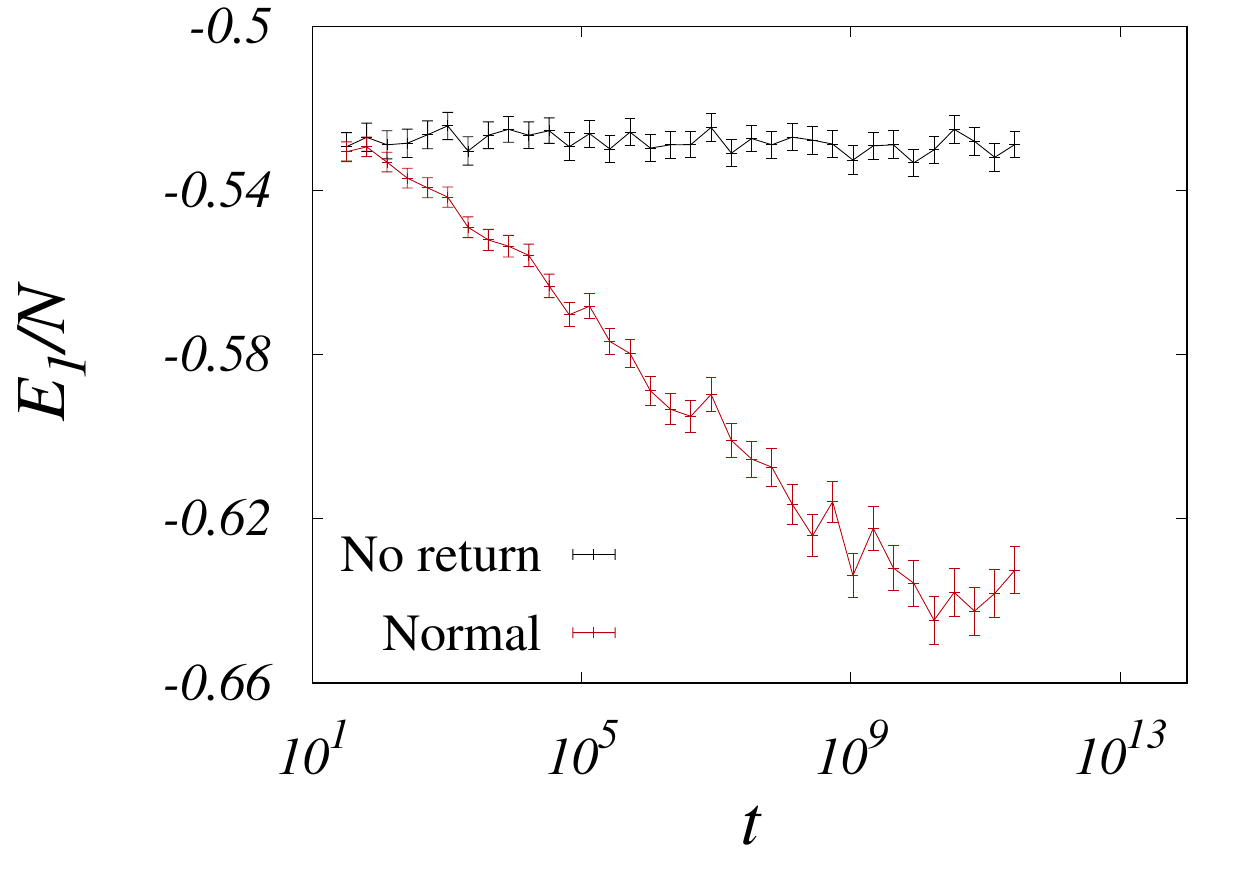}
\includegraphics[width=0.49\textwidth]{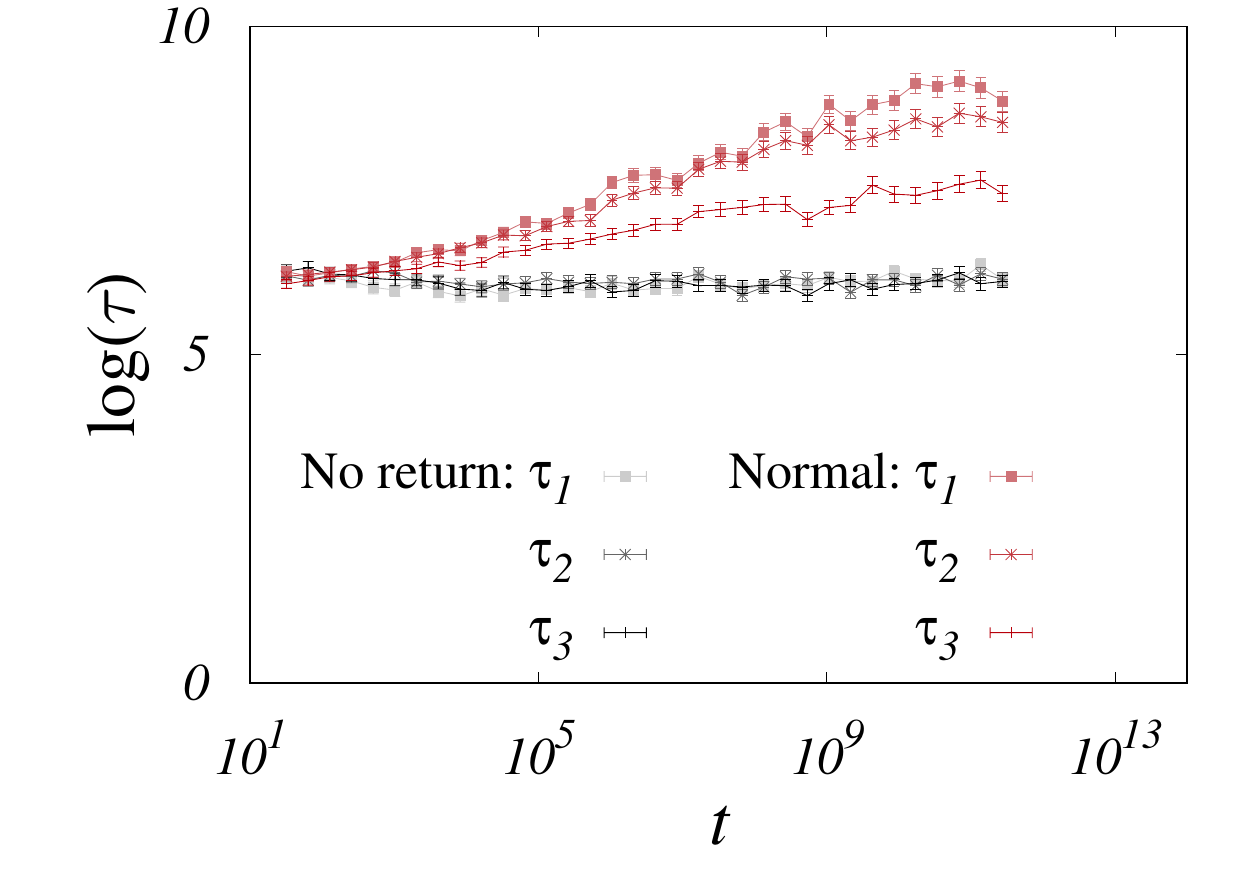}
 \caption{Left: 
 comparison between $E_1$ in the \ac{rem}, and its counterpart $E_1^\nr$ that does not take in account returns to
 the just-visited basin. While $E_1$ is logarithmic in time, $E_1^\nr$ is constant, recovering the \ac{tm} prediction.
 Right:
 comparison between the trapping times $\tau_1,\tau_2$ and $\tau_3$, and their non-returning counterparts
 $\tau_{1}^\nr,\tau_{2}^\nr$ and $\tau_{3}^\nr$ in the \ac{rem}. Also in this case the curves become constant and collapse on each other
 once the returns are ignored.
 In both plots we have $N=20$, $T=0.25$.
}
 \label{fig:noret}
\end{figure}

The non-renewal features of the previous section can be explained in a simple way in terms of  \emph{returns} to the same basin. 
While in the \ac{tm} the phase space connectivity graph is fully connected, in the \ac{rem} it is a hypercube, since a movement is
obtained by flipping one of the $N$ spins.\\
More explicitly, in a \ac{tm}, from each configuration the system can reach any of the $2^N$ states. On the contrary, 
in the \ac{rem} only $N$ configurations are connected to each site, and of those only few (or one) of them are typically visited.
The smaller amount of available paths in configuration space conveys that there is a finite probability that, when the system leaves 
a valley, the next visited basin will be again that same one. What happens is that when the system manages to jump out of a deep 
configuration it lands in a new configuration close to the threshold level. However, the dynamics and the configuration space at the threshold level is complicated (see later for a discussion) and before being able 
to wander far from the initial configuration, the system repeatedly visits it. If the system is
in a very deep basin, it is likely that the next basin be very deep too, since it is likely that the same basin be visited again. \\
Let us now confirm these statements by simulation results. 
We call $S$ the number of spins that are different between each basin and the following one.
$S=0$ means that the same basin was visited twice in a row.
\begin{figure}[!htb]
\includegraphics[width=\textwidth]{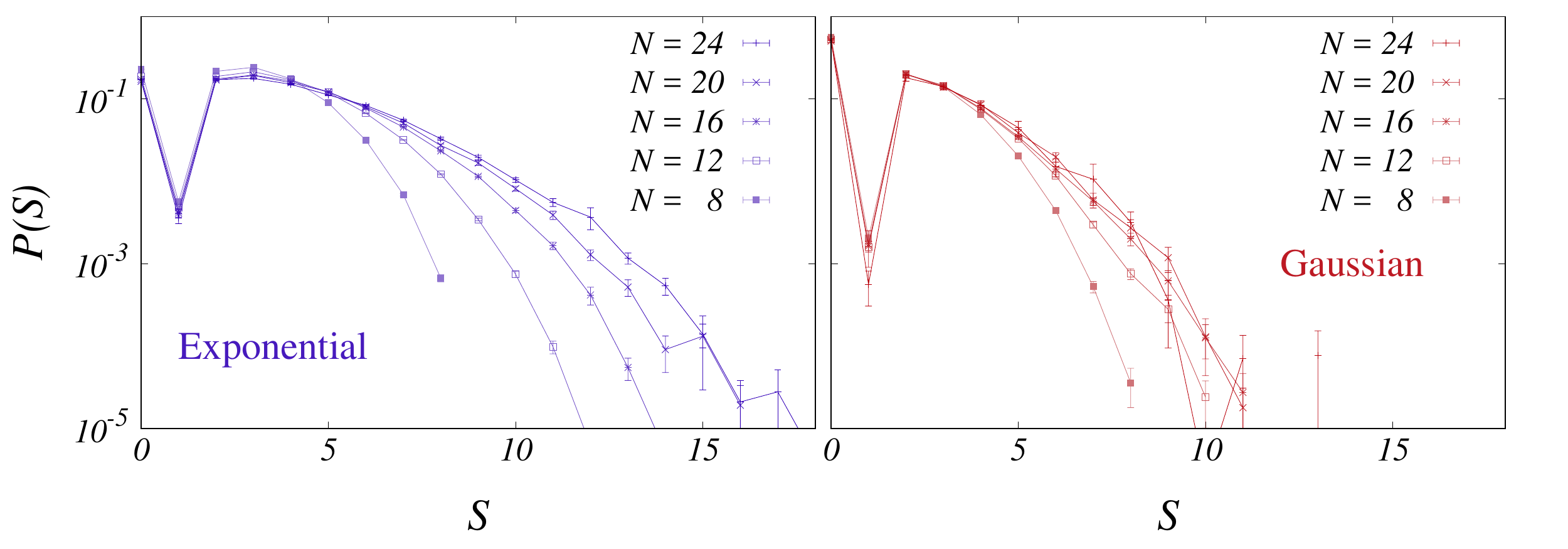}
 \caption{$P(S)$ in the Exponential (left) and Gaussian (right) \ac{rem}, at $T=0.25$. A large fraction of times
 the system comes back to a just visited basin ($S=0$). This fraction does not show a visible dependence on $N$: already 
 for $N=20$ the size dependence seems to have saturated.
 }
\label{fig:pS12}
\end{figure}
In figure \ref{fig:pS12} we show that $S=0$ a finite fraction of the times. 
{
Note that $P(S)$ is an indicator of the basin size and the distance between basins. 
When increasing $N$, the $P(S)$ seems to converge to a limiting function, independent of $N$. In other words,
for large enough sizes $P(S)$ does not depend on $N$. Since the typical basin size is 1 by construction (in the REM two configurations differing by one spin-flip have uncorrelated energies), this indicates
that the distance between basins is also $\mathcal{O}(1)$. 
A second indication of this is found in Sec.\ref{sec:trapping-times},
where the times spent on the barrier do not grow with $N$.\\
Finally, let us point out that one would expect that the distance between two different basins be at least $2$, since
there should be at least one barrier configuration between them. The fact that $P(S=1)$ is positive, though small, 
descends from the fact that occasionally neighbouring configurations with $E<\eth$ may be counted as two separate basins at distance $1$, because of our dynamic definition
of basins. This happens when a transition to $E>\eth$ was accepted after visiting one, and before visiting the other. 
} \\
We can quantify the phenomenon of immediate returns in the same basin by computing the probability that the system falls twice in a row in the same basin. We will focus on the simplest situation,
that we call \emph{immediate returns}, in which the system returns to a basin immediately after having left it. This particular case provides a lower bound to the frequency of returns, and can thus
be used to state that returns occur when $N\to\infty$.\\
Until now, we have studied the structure of the configurations dynamically connected to a given $C_0$, knowing that this structure does not depend on the chosen $C_0$, due to the independence of the energies from the configurations in the \ac{erem} and \ac{rem}.
For the computation of the probability of immediate returns we will assume that the system just left a basin. In other words we will consider the case of a configuration $C_0$ on the barrier ($E_0> E_{th}$) 
that is just one spin flip apart from a configuration $C_1$ that was deep down in the energy landscape ($E_1<E_{th}$).  Apart from this, the other $N-1$ configurations $C_2,\dots,C_N$ dynamically connected to $C_0$ follow the standard statistics described by \eqref{eq:kth-deepest-rem} and \eqref{eq:kth-deepest-erem}.
By construction, $E_1\ll\eth\leq E_0$ (see figure \ref{fig:diagram}).\\
\\
We now estimate the probability of \emph{immediate returns} by simple arguments. 
We call $\Pacci$ the probability of accepting a move from $C_0$ to $C_i$. According
to the Metropolis rule,
\begin{equation}
 \Pacci = \min[1,e^{\beta(E_0-E_i)}]\,.
\end{equation}
In a sequentially updated dynamics, the system returns straight back to $C_1$ when the other $N-1$ attempted updates, leading to $C_2,\ldots,C_N$, are rejected. 
Since each move is accepted independently, the probability of rejecting all the $N-1$ moves is
\begin{equation}\label{eq:pret}
 \Prej=\prod_{i=2}^{N}(1-\Pacci)\,.
\end{equation}
Once any other move has been rejected we are sure that the system will go back to $C_1$, since $E_1<E_0$.
Hence, to evaluate the probability of an immediate return, we need to estimate the probability of rejecting all the other moves, paying attention to the 
energy structure of the configurations $C_2,\dots,C_N$ (there is no need to consider a sequential dynamics or that all moves are attempted before $C_1$, 
since the only thing that matters is the scaling of $\Prej$ when $O(N)$ attempts are considered). \\
Since the energy differences between $E_0$ and $E_i$ scale in a different way in the \ac{rem} and \ac{erem}, we consider the two cases separately.
\begin{itemize}
\item {\bf \acs{rem}}. The energy $E_{0}$ is by definition typically of the same order (but slightly larger than) $E_{th}$.
The neighbours of $C_0$ do not coincide with the configurations surrounding the initial configuration $C_1$, the one deep down in the energy landscape (figure \ref{fig:diagram}). 
Thus, in order to estimate the landscape around 
$C_0$, one has to consider the minimal energies among the $N-1$ other neighbours of $C_0$. These follow eq.~\eqref{eq:kth-deepest-rem}, since considering $N$ or $(N-1)$ 
neighbours leads to subleading corrections only. As a consequence, we obtain that the typical difference between 
$E_0$ and the energy of its lowest neighbouring configuration (excluding $C_1$) scales as 
$\delta^{(2)}\sim\sqrt{\frac{N}{\log N}}$ in the large $N$ limit. The probability distribution of this difference can be 
obtained by using results from extreme value statistics, but for our purposes it suffices to say that, with finite probability $P^\mathrm{up}<1$,
the lowest neighbour of $C_0$ (excluding $C_1$) has a higher energy, of order
$E_0+\mathcal{O}({\sqrt{\frac{N}{\log N}}})$. In this case, since all other neighbours are even higher in the energy landscape, 
the probability of accepting a move to one of these neighbours is exponentially small in $\sqrt{\frac{N}{\log N}}$; 
therefore even after $O(N)$ attempts (the number of neighbours) it remains negligible, i.e. the probability of returning goes to one at large $N$. Note that with probability $1-P^\mathrm{up}$, instead, the 
lowest neighbour of $C_0$ (excluding $C_1$) has an energy of order $E_0-\mathcal{O}(\sqrt{\frac{N}{\log N}})$.
In these cases there is no immediate return since the system directly goes to this neighbour if this move is attempted before the one towards $C_1$; 
depending on the neighbours of this neighbour, the system could then be forced to return to $C_0$ instead of exploring further the landscape. 
\item {\bf \acs{erem}}. In the \ac{erem} the computation needs slightly more care, because the separation $\delta^{(2)}$ between the lowest energy levels is of order one. 
We will keep the reasoning simple and present just a rough (but essentially correct) argument. We call again $P^\mathrm{up}$ the finite probability less than one 
that the lowest neighbour of $C_0$ (excluding $C_1$)
has a higher energy of order $\delta^{(2)}\sim \mathcal{O}(1)$. 
Let's say that in this case we divide the neighbours of $C_0$ (excluding $C_1$) in two groups: $n$ which are the lowest 
in the landscape and have therefore positive energy differences of the order one and $N-n-1$ which are typical configurations 
and have typical energies, hence a difference of the order $\alpha \log N$.  Within this construction, $n$ is finite. 
We can then estimate that
\begin{equation}
\Prej>\left(1-e^{-\beta\alpha}\right)^n\left(1-e^{-\beta\alpha\log(N)}\right)^{N-1-n} \ ,
\end{equation}
which in the large $N$ limit goes like 
\begin{equation}
\Prej>\left(1-e^{-\beta\alpha}\right)^ne^{-N^{1-\beta\alpha}} \ .
\end{equation}
In the high temperature regime, $\beta<1/\alpha$ (where there is no aging) the lower bound goes to zero, while in the low temperature regime, $\beta>1/\alpha$, 
(where there is aging) the lower bound remains finite and equal to $\left(1-e^{-\beta\alpha}\right)^n$. This implies that with finite probability the system returns back to the initial basin.
A refined argument can be worked out and leads to the same conclusion, which is actually expected from the thermodynamics of the EREM: 
for $\beta<\beta_c$ the system is sucked up by configuration at zero intensive energies whereas for $\beta>\beta_c$ the system can return back to $C_1$, 
but it does so with probability less than one since it can 
accept a move to a neighbour which has an energy difference of order one.\\
In cases where the lowest neighbour of $C_0$ (excluding $C_1$)
has a lower energy of order $\delta^{(2)}\sim O(1)$ an analogous scenario holds, except that now if the 
move to the neighbour with lower energy is attempted before the one to $C_1$ the system does not 
effectuate an immediate return with probability one (but it can return with finite probability in a finite number of steps).    
\end{itemize}
Overall we have found that the dynamics after the jump out from $C_1$ is complicated and cannot be described 
as trap-like: it actually corresponds to a complex wandering on configurations at the threshold level and above. 
This explains the numerical results shown at the beginning of this section. We can further confirm that the returns 
are at the origin of the numerical findings by defining $E_1,\tau_1,\tau_2$ and $\tau_3$ in a different way, that suppresses the effect from the 
returns. We define $E_1^{\nr}$ as the energy of the first basin encountered, after $t$, with $S>0$, i.e. imposing that the basin is not the initial one. 
The definition of $\tau_1^{\nr}$ is analogous. Regarding $\tau_2^{\nr}$ and $\tau_3^{\nr}$, we prescribe that all the recorded basins since $t$ be new ones.\\
In figure \ref{fig:noret} we show that both $E_1^{\nr}$ and $\tau_1^{\nr}, \tau_2^{\nr}$ and $\tau_3^{\nr}$ fulfill the 
trap prediction now that the returns are not being recorded.
Note, however, that although this indicates trap-like dynamics, we do not expect to find renewal dynamics and 
hence all trap predictions to hold on the time-scale on which the system performs a complex wandering at the 
threshold level and above. This happens only on much larger time-scales as discussed and showed in the next section.

\section{Trap-like dynamics on long time-scales}
 From the previous analysis we know that when the system is in a basin whose energy is extensively below zero, i.e. $\lim_{N\rightarrow \infty}E/N<0$, it takes a time $\tau_{J}$ which diverges as $e^{-\beta E}$ to jump out of it. The dynamics that follows this first jump consists in a wandering on configurations at and above the threshold level and in returns to the initial basin.  This exploration takes place on times
that differ from $\tau_{J}$ by a multiplicative factor which vanishes at exponential leading order in $N$.
For example, for the REM the barriers that the system has to overcome at the threshold level are of the order $\sqrt{N/\log N}$
therefore, even taking into account returns to the initial basin, the time-scale $\tau$ to wander at the threshold level
is still $e^{-\beta E}$ at exponential leading order: $(\log \tau)/N=-\beta E/N\simeq -\beta E/N +\sqrt{1/N\log N}$
for $N$ large. \\
As discussed previously, the dynamical processes following the jump out of the basins
are not trap-like and, instead, correspond to a quite complex evolution in configuration space.
However, if one observes motion on the time-scale  $e^{-\beta E}$, which packs together
all times $\tau$ that differ for a multiplicative factor vanishing
at exponential leading order in $N$, then the wandering at the threshold level becomes a very fast process.
 On this time-scale, the system does many back and forth motion and explores new configurations.
 In particular, it visits a number of new configurations that diverges as a power law with $N$; this
 should be enough to loose memory and, hence, effectively move to another uncorrelated basin.
 Analogously, although there are correlations in the energies sampled during the dynamics at the threshold level, they do not extend on timescales and hence on a number of different visited configurations which diverge exponentially in $N$.  \\
 All that suggests that once dynamics is coarse-grained on time-scales exponentially diverging with $N$
 an effective trap-like description should emerge, in particular we expect the fundamental pillars of trap-like dynamics to hold:
\begin{enumerate}
\item At time $t$ the lowest intensive energy $e(t)$ sampled is given by the equation $t \sim e^{-N\beta e(t)}$.
\item The number of basins $\mathcal N(t)$ visited until time $t$ is given at the exponential accuracy in $N$ by the number of times one has to draw Gaussian random energies in order to get a minimum of the order of $e(t)$: $\mathcal N(t)\sim e^{Ne(t)^2}$.
\item  Thermal equilibrium sets in for all configurations from energy zero to energy $e(t)$, i.e. at time $t$ one gets an effective Boltzmann distribution cut at energies lower than $e(t)$.
\end{enumerate}
These features imply the emergence of trap-like aging dynamics on exponentially diverging time-scales, as indeed very recently proven in \cite{gayrard:17b}.  \\
In the following, we confirm this result through numerical simulations. As we shall show, important finite size-time effects 
correct the trap-like predictions. Low temperatures make the dynamics so slow that it is very difficult to probe 
the regime in which energies are extensively below $\eth$. At the same time, important finite size effects are observed for temperatures close to $T_c$. All that is particularly cumbersome in the REM, where the fact that trap-like aging emerges on an infinite spectrum of exponentially diverging time scales makes the analysis even more difficult.
\\
Note that above we have stated the trap-like properties in terms of basins and not configurations. As discussed in the previous sections a basin typically contains just one configuration. For observables that probe typical behaviour
as for example the aging function  
considering basins or configurations is equivalent; however for observables for which rare effects are enhanced 
this might be incorrect. For instance, the number of jumps done until time $t$ behaves differently if one considers basins or configurations. In the former case the usual trap prediction holds, whereas in the latter the back and forth 
motions within rare basins formed by more than one configuration bias completely the statistics which, depending on the regimes, can be different from the trap prediction; see \cite{gayrard:17b} for a detailed discussion and results. 
Moreover, also from the numerical point of view, considering basins or configurations is not equivalent. As we shall show, the former 
leads to smaller asymptotic corrections and is therefore preferable.
\subsection{Logarithmic decrease of the energy}\label{sec:elogt}
During aging, the trap prediction is that the intensive energy decreases logarithmically in time as $-T(\log t)/N$, see equation \eqref{eq:et}.
\begin{figure}[!tb]
\includegraphics[width=\textwidth]{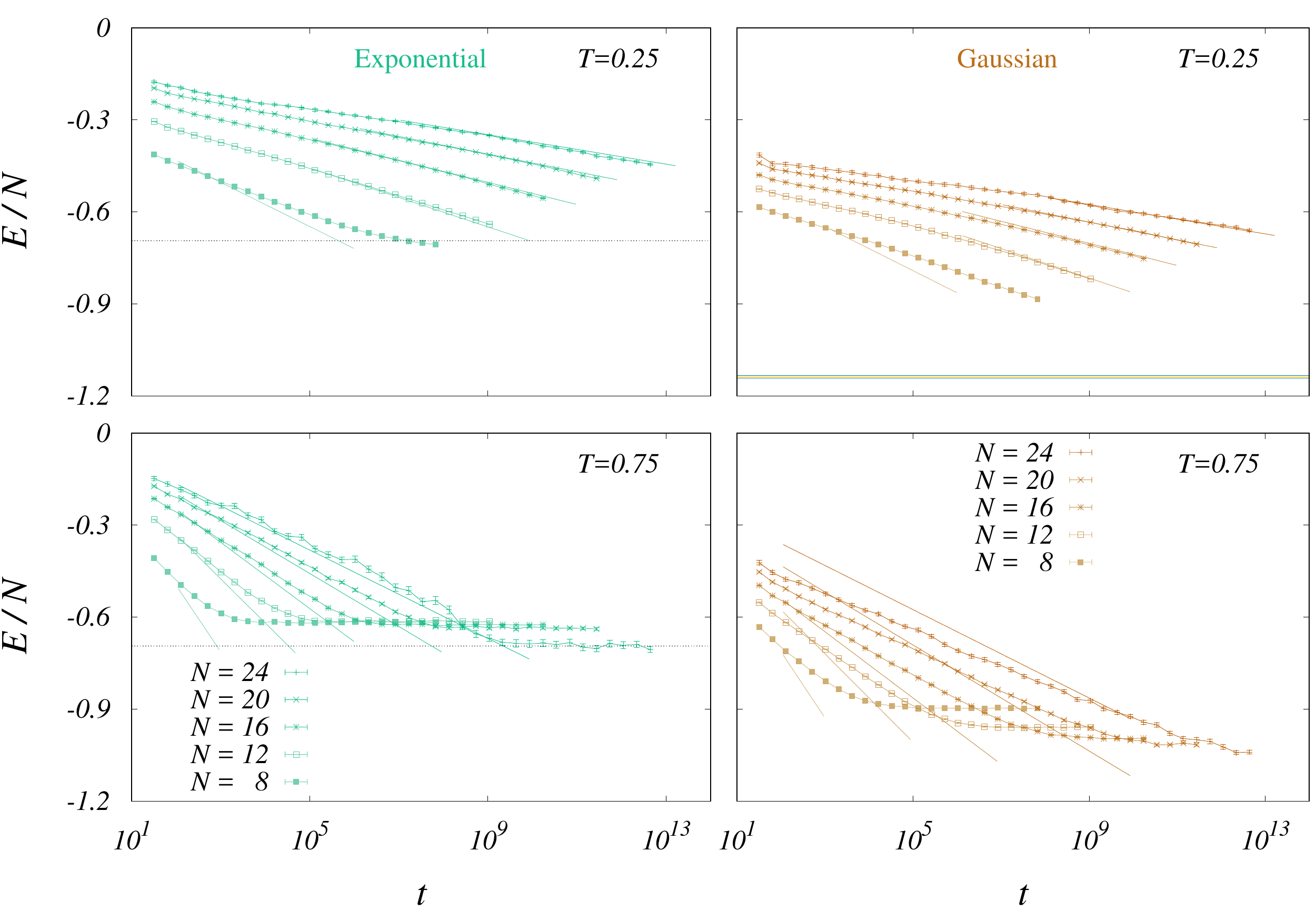}
 \caption{$E(t)$ in the \ac{erem} (left) and in the \ac{rem} (right). The temperature is $T=0.25$ on top, and $T=0.75$ on the bottom. The solid lines
 represent equation \eqref{eq:et}. The dotted horizontal lines on the left side indicate the ground state energy. When the curve $E(t)$ reaches
 its equilibrium value, the system has thermalized, and it is no more in the aging regime. The solid horizontal lines in the top-right 
 plot are the predictions of the equilibrium free energy according to reference \cite{derrida:81}, which takes into account logarithmic corrections. 
 It was not possible to plot the same predictions for $T=0.75$, because the logarithmic order was not enough to obtain reasonable predictions.
}
\label{fig:etbar}
 \end{figure}
Figure \ref{fig:etbar} shows that, at $T=0.25$, for the largest system sizes this is fairly reproduced both for \ac{erem} and \ac{rem}; the deviation from the logarithmic behaviour is mainly due to equilibration at long-times. 
The result is not as good when $T=0.75$. We attribute this to the emergence of extremely strong finite-size effects as $T$ approaches $\Tc$. 
{This explanation is confirmed by the fact that, despite being 
very big for $T=0.75$ and smaller for $T=0.25$, deviations between \ac{tm} 
predictions and observations decrease as $N$ grows in both cases.}
As a matter of fact, the finite-size corrections to the free energy
diverge as $T\to\Tc$ \cite{derrida:81}, so the closer the system is to the critical temperature, the larger it needs to be.
\subsection{Trapping times}\label{sec:trapping-times}
\begin{figure}[!tb]
\includegraphics[width=\textwidth]{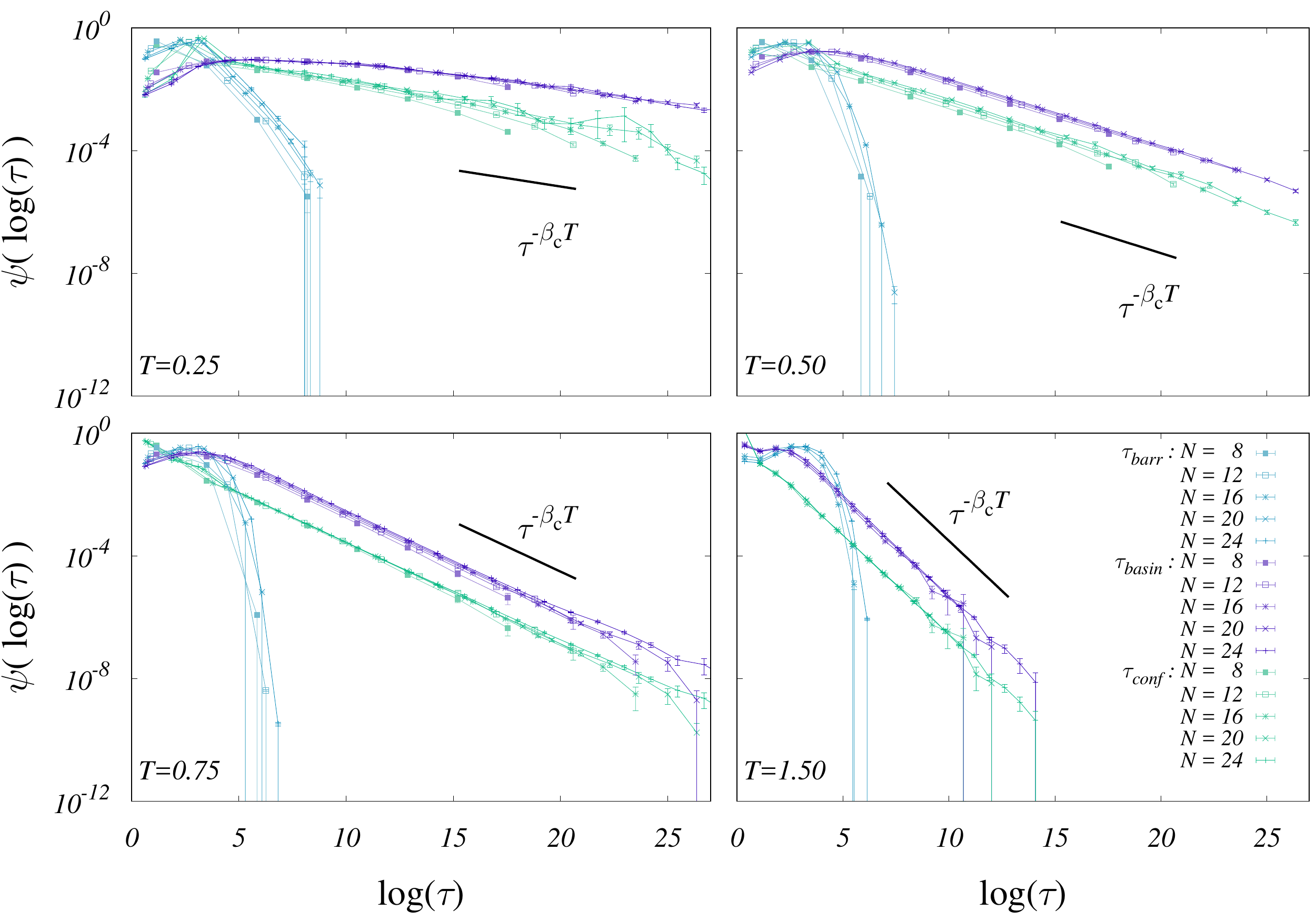}
\caption{The distribution of the logarithm of the trapping times $\psiexp(\log(\tau))\propto\tau^{-T/\alpha}$ with $\beta_c=1/\alpha$ in the \ac{erem} for $T=0.25, 0.50, 0.75, 1.50$.
The red data are the times $\tau_\mathrm{conf}$ before each configuration is abandoned,
the blue data represent the time $\tau_\mathrm{basin}$ spent in each basin, and the green data are the times spent on the barriers.
The black line is the trap prediction \eqref{eq:ptau-expo}. The exponent is different because we are plotting $\psiexp$
as a function of $\log(\tau)$.}
\label{fig:pTau-exp}
\end{figure}

\begin{figure}[!htb]
\includegraphics[width=\textwidth]{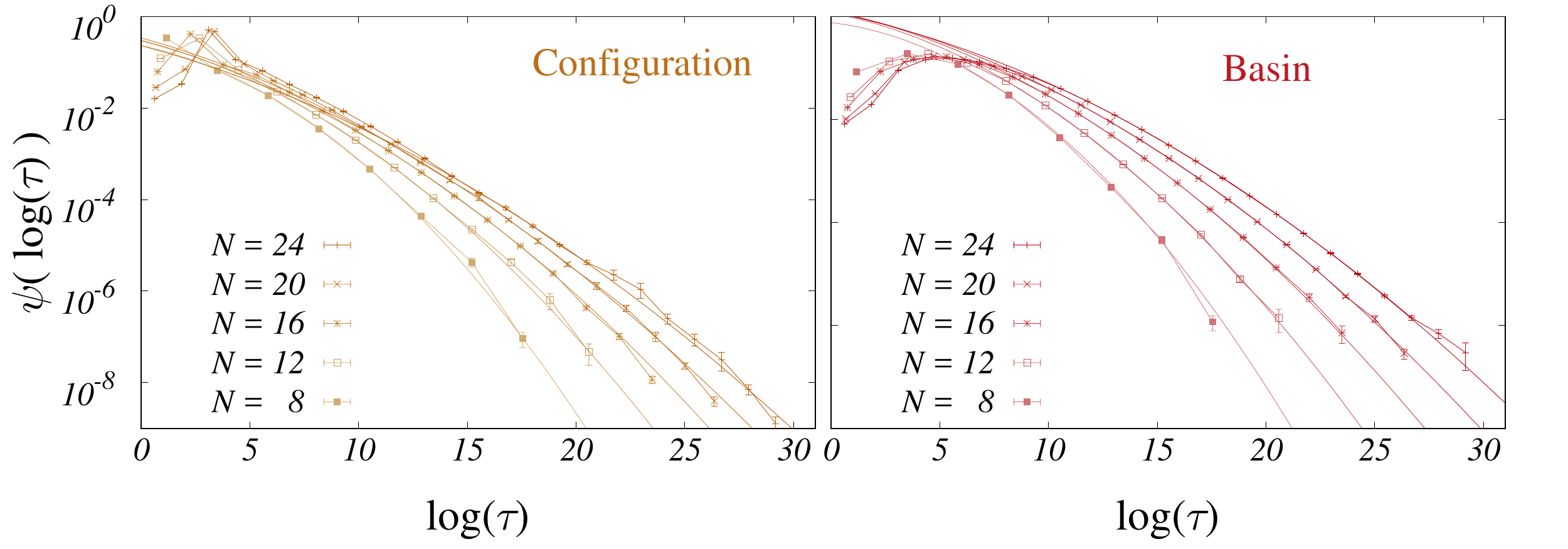}
 \caption{Distribution of trapping times $\psigauss(\tau)$ in the Gaussian \ac{rem}, for $T=0.75$. Configuration trapping times on the left, and basins on the right.
 The curves represent fits to the function $f(\log(\tau))=A\exp(-\frac{T^2}{2N}\log(\frac{\tau}{\tau_0})^2)$, which is the long-time prediction
 of $\psigauss(\tau)$ [equation \eqref{eq:ptau-gauss}].
 Fits exclude the last point, because we expect that the finiteness of our simulations can bias on the sampling of the longest times.
 }
 \label{fig:pTau-gauss}
\end{figure}

The hallmark of a trap-like dynamics is the power-law distribution of the trapping times $\psiexp(\tau)$. 
In the \ac{erem} case it follows the power-law behaviour of equation \eqref{eq:ptau-expo}. 
At $T=0.50,0.75,1.50$ the prediction is beautifully satisfied both for the basin and for the configuration trapping times, even for $T>\Tc$ (figure \ref{fig:pTau-exp}).
This is not the case for $T=0.25$. At this temperature dynamics is so slow that the aging regime cannot be explored in extent, one needs longer 
times to enter the trap-like regime in which the typical configuration has $E<\eth$. Anyhow, the data are compatible with the trap prediction
both in the exponent and in the fact that basin and configuration observables behave in the same way.
As anticipated above, some tuning of the parameters (temperature, size of the system) is necessary in order to avoid big finite time-size corrections. 
\\
In the \ac{rem} the trap prediction for $\psigauss(\tau)$ follows equation \eqref{eq:ptau-gauss}, which is not a power law.
\footnote{Only on a given exponentially diverging time-scale the usual power law is recovered as we recalled in Section 2.}
As found for the \ac{erem}, at $T=0.25$ the simulations require too long times and it is not possible to compare $\psigauss$ to the prediction.
On the other side, at $T=0.75$, the data fit very well both for the basins and for the configurations (figure \ref{fig:pTau-gauss}).

{
\paragraph{A note on the consistency of the results at different temperatures}
Since all temperatures under $T_\mathrm{c}$ are equivalent when
one seeks the qualitative behaviour of the model, in principle showing
results only for one temperature would be enough. There is a reason
why one should focus on more than one temperature, and it is that
there are two sources of quantitative (not qualitative) changes which
depend on $T$, and we can only choose to rule out one or the other in a single simulation:
\begin{itemize}
 \item[(a)] When $\Tc$ is approached, the finite-size effects on the free and
equilibrium energy are larger \cite{derrida:81}. 
Thus, all other things being equal, close to Tc 
one has to consider larger system sizes. 
 \item[(b)] Many of the dynamical behaviors we are interested in only appear on exponentially long time 
scales, $t>>\exp(a(T)N)$, with $a(T)$ growing as $T$ is lowered. As a
consequence, the temperature has to be high enough otherwise 
interesting phenomena emerge on prohibitively large time-scales. 
\end{itemize}
Given the competition between the two effects, one happening at low and the other at high temperature, 
we simulated several temperatures, in order to be able to be in the correct regime for each
observable.
}

\subsection{Aging functions} \label{sec:aging}
As discussed in section~\ref{se:tm}, all the relevant information of the correlation functions for trap dynamics is contained in the aging function $\Pi(\tw,t)$, which represents the probability 
of not changing trap. To study the trap behaviour of the aging dynamics in the \ac{rem} and \ac{erem}, we observed the behaviour of this function by considering either 
basins and configurations for the sake of comparison (as done in~\cite{cammarota:15}). We then defined the probability of not changing basin $\Pi_\basin(\tw,t)$ or configuration 
$\Pi_\conf(\tw,t)$ between a time $\tw$ and $t=\tw(1+w)$ $(w>0)$. 
\subsubsection{Aging functions in the EREM	}
\begin{figure}[!t]
\centering
\includegraphics[width=\textwidth]{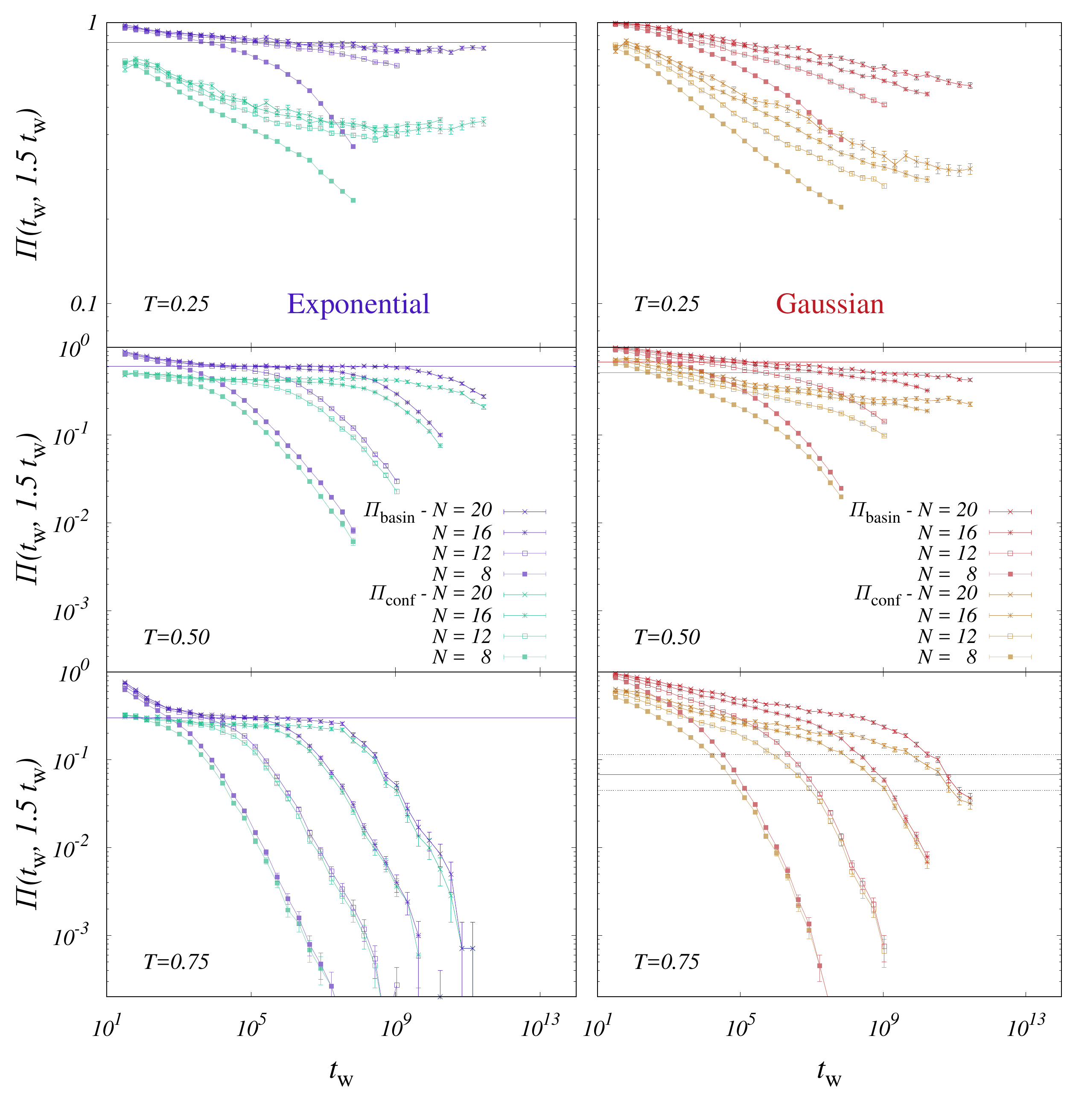}
 \caption{Aging functions \ac{rem} for $w=0.5$. From top to bottom: $T=0.25, 0.50, 0.75<\Tc$. On the left
 we depict the Exponential, on the right the Gaussian \ac{rem}. We show the basin and configuration aging functions, for different system sizes.
 Note the different scale for $\Pi$ in the two top sets.
 In the left plots, the blue horizontal lines represent the infinite-time limit of the aging function
 for an infinitely large system [recall equation \eqref{eq:pi}].
 The horizontal lines in the center-right plot show the effective value of the effective infinite-time prediction
 for the aging functions (see main text). The top one is calculated on the basin's $\psigauss(\tau)$,
 the lower one is on the configuration $\psigauss(\tau)$.
 The horizontal lines on the bottom-right plot represent a confidence interval for the same quantity, averaged between
 basins and configurations.
 }
 \label{fig:Pi}
\end{figure}

In the large-time limit of aging dynamics of \ac{tm}s, $\Pi(\tw,\tw(1+w))$ converges to a value $H_x(w)$, with $x=T/\alpha$, that can be computed analytically \cite{bouchaud:95,benarous:06}.
In figure \ref{fig:Pi} we show the aging functions $\Pi_\basin(\tw,t)$ for different sizes $N$ and different temperatures in the \ac{erem} (left) and \ac{rem} (right).\\
We can distinguish two regimes, an initial plateau regime corresponding to the aging regime (associated to 
$t_w\rightarrow \infty$ after $N\rightarrow \infty$) 
and a long-time decay once the system thermalised (associated to 
$t_w\rightarrow \infty$ before $N\rightarrow \infty$). 
The plateau becomes more and more pronounced as the system size increases, as expected.
In the plateau regime, the trap prediction given by $H_{x}(w)$ applies well to the basin aging functions 
and less well to the configuration aging function, indicating that activated dynamics in the \ac{erem} is effectively trap-like as long as we focus on basins.
We interpret the discrepancy between the \ac{tm} prediction and the configuration aging function as 
a finite time effect: the lower the temperature, the larger the time needed to enter in the trap-like
dynamics regime where the energy is intensively lower than $\eth$ and a basin is really formed 
by just one configuration. Accordingly, we should observe for very large system sizes (for which the 
plateau of the aging function extends to very large time scales) a slow upwards drift of the plateau 
of the configuration aging function towards the plateau of the basin aging function. A hint of this 
upward drift seems to appear in the long time limit of aging function obtained at the lowest observed 
temperature $T=0.25$ and for the largest system size considered in figure \ref{fig:Pi}.
\subsubsection{Aging functions in the REM}
In the case of the \ac{rem}, the interpretation of the data is less straightforward, as it is seen 
in figure \ref{fig:Pi} (right), because the plateau regime is not well-defined. This is due to the fact that 
in the \ac{rem} the trap-like dynamics is visible only if the exponentially diverging time-scales are well separated. 
This would only happen in the right scaling limit of infinitely large systems and times.
We can try nevertheless to extract an effective aging function to compare with.
Equation \eqref{eq:ptau-gauss} can be seen as a power law with an exponent that depends on the simulation time: according to the length
of the run, one can extract an effective exponent $x_\mathrm{eff}(t)=\frac{T}{T_c^{_\mathrm{eff}}(t)}$. \\
Two equivalent ways can be used to extract it.
The first method is to fit to a power law the final part of the curve $\psi(\log(\tau))$. By comparison with equation \eqref{eq:ptau-expo}, $x_\mathrm{eff}(t)$
is the exponent of the curve. A second way is to use the relation, valid in the asymptotic limit,  $x_\mathrm{eff}(t)=-\frac{TE(t)}{N}$ that only requires the knowledge of the average energy $E(t)$ reached in the simulations at time $t$.\\
The two methods of extracting the effective asymptotic values of the aging functions give compatible values,
that are plotted in figure \ref{fig:Pi} (right){, and show that numerical results are not incompatible with a trap behaviour}. 
Further, as stated in section \ref{sec:trapping-times}, at the lowest temperature, $T=0.25$, the simulation time was too small to sample
well the phase space. As a consequence, we could not extract an effective exponent $x_\mathrm{eff}$.
Clearly much larger size and times would be needed to enter the asymptotic regime. 
{Remarkably, a recent numerical study of generalised trap models \cite{cammarota:17} has shown that 
the very same problem also arises in simpler models with a gaussian distribution of energies. 
The analysis developed in \cite{cammarota:17} could be applied to the REM case 
in order to find more stringent evidences  of trap dynamics.}
\\
Finally in agreement with the results for the 
\ac{erem}, we find also in this case a more pronounced difference at low temperature between basin and configuration descriptions.

\section{Conclusions}
We have shown how the main features of TM-like dynamics emerge effectively on exponentially diverging time-scales 
in the case of the simplest mean-field glassy models (\ac{rem} and \ac{erem}). 
Despite motion in configuration space is complex on intermediate time-scales, in particular at the threshold level and above, 
with frequent back and forth movements in configuration space, the very large separation of time-scales washes out correlations induced by the exploration of the landscape, and hence memory. 
As a result, by coarse-graining dynamical evolution in terms of basins, a full \ac{tm} description of the dynamics holds,
in agreement with very recent rigorous results \cite{gayrard:17b}. Identifying traps with basins instead of 
configurations is, as we have seen, important both numerically, to reduce pre-asymptotic corrections, and
theoretically (see \cite{gayrard:17b}). Only in this way a complete \ac{tm} description can be reached. This will be even more crucial in
studies of more complex and rich models, for which our results provide guidelines. \\[1ex]
The natural next step is to study the 
p-spin model, that in the large $p$ limit converges to the \ac{rem} \cite{derrida:80,gross:84}. In our study we found
that in the activated dynamics regime the system has always to reach and wander at the threshold level in order to escape from a basin and loose memory of the past. Whether this holds in general is a very interesting question
which would shed light on whether an effective TM description is valid more generally. An interesting possibility is that to escape from a deep minimum the system only needs to reach an energy level which is lower than the threshold, which is identified as the lowest energy reached during aging dynamics on time-scales of order one. This would be an important first piece of information to understand activated dynamics in realistic system such as super-cooled liquids whose energy landscape displays strong similarities 
with the one of mean-field glassy systems \cite{castellani:05}. \\
These questions can be addressed applying our approach to 
$p$-spin or more general models. Work is in progress in this direction \cite{inpreparation}.  

\section*{Acknowledgements}
We benefit from a lot of interesting and useful discussions with G. Ben Arous about \ac{rem}, \acp{tm} and rigorous results 
on aging dynamics.  We also thank Valerio Astuti, Alain Billoire, Ivailo Hartarsky and Enzo Marinari for interesting discussions.

\paragraph{Funding information}
This work was funded by the Simons Foundation for the collaboration ``Cracking the Glass Problem" (No. 454935 to G. Biroli), and 
the ERC grant NPRGGLASS (No. 279950). 
M.B.-J was partially supported through Grant No. FIS2015-65078-C2-1-P, jointly funded by MINECO (Spain) and FEDER (European Union).

\begin{appendix}

\end{appendix}



  \bibliographystyle{unsrt} 
\bibliography{./marco.bib}

\nolinenumbers

\end{document}